\UseRawInputEncoding
\documentclass[aps,showpacs,epsf,two column,pra]{revtex4}
\usepackage{amssymb}
\usepackage{amsmath, bm}
\usepackage{graphicx,psfrag}
\usepackage{braket}
\usepackage{float}
\usepackage{tikz,siunitx}
\usetikzlibrary{shapes.geometric,shapes.symbols}

\usepackage{tikz}
\usepackage{epstopdf}

\usepackage{pgfplots}
\makeatletter

\newcommand{\Rmnum}[1]{\expandafter\@slowromancap\romannumeral #1@}
\newcommand{\colorcaption}[2][]{
  \begingroup
  \renewcommand{\@caption@fignum@sep}{ (color online). }
  \caption[#1]{#2}
  \endgroup
  }
\makeatother
\usepackage[colorlinks=true, citecolor=blue, urlcolor = blue, linkcolor= red, bookmarks=true]{hyperref}
\begin{document}
\newcommand{\markerone}{\raisebox{0.5pt}{\tikz{\node[draw,scale=0.4,circle,fill=black!20!blue](){};}}}
\newcommand{\markertwo}{\raisebox{0pt}{\tikz{\node[draw,scale=0.3,regular polygon, regular polygon sides=3,fill=black!45!red,rotate=180](){};}}}
\newcommand{\markerthree}{\raisebox{0.5pt}{\tikz{\node[draw,scale=0.4,regular polygon, regular polygon sides=3,fill=black!10!red,rotate=0](){};}}}
\newcommand{\markerfour}{\raisebox{0.5pt}{\tikz{\node[draw,scale=0.4,regular polygon, regular polygon sides=4,fill=red](){};}}}
\newcommand{\markerfive}{\raisebox{0pt}{\tikz{\node[draw,scale=0.4,diamond,fill=black!10!red](){};}}}
\newcommand{\markersix}{\raisebox{0.6pt}{\tikz{\node[draw,scale=0.3,circle,fill=black!100!](){};}}}
\def \beq{\begin{equation}}
\def \eeq{\end{equation}}
\def \bse{\begin{subequations}}
\def \ese{\end{subequations}}
\def \bea{\begin{eqnarray}}
\def \eea{\end{eqnarray}}
\def \bem{\begin{displaymath}}
\def \eem{\end{displaymath}}
\def \bem{\begin{bmatrix}}
\def \eem{\end{bmatrix}}
\def \Ps{\hat{\Psi}(\boldsymbol{r})}
\def \Pds{\hat{\Psi}^{\dagger}(\boldsymbol{r})}
\def \i{{\int}d^2{\bf r}}
\def \bl{\bar{\boldsymbol{l}}}
\def \c{\hat{c}_{n,m}}
\def \cp{\hat{c}_{n',m'}}
\def \cd{\hat{c}_{n,m}^{\dagger}}
\def \cdp{\hat{c}_{n',m'}^{\dagger}}
\def \bb{\bibitem}
\def \nn{\nonumber}

\def \bs{\boldsymbol}
\def \hkx{\hat{k}_{x}}
\def \hky{\hat{k}_{y}}
\def \bq{\bar{q_{y}}}

\def \bc{\begin{center}}
\def \ec{\end{center}}

\title{\textbf{Bogoliubov spectrum and the dynamic structure factor in a quasi-two-dimensional spin-orbit coupled BEC }}
\author{Inderpreet Kaur and Sankalpa Ghosh}
\affiliation{Department of Physics, Indian Institute of Technology Delhi, New Delhi-110016, India}

\begin{abstract}
We compute the the Bogoliubov-de-Gennes excitation spectrum in a trapped two-component spin-orbit-coupled (SOC) Bose-Einstein condensate (BEC) in quasi-two-dimensions as a function of linear and angular momentum and analyse them. The excitation spectrum exhibits a minima-like feature at finite momentum for the immiscible SOC-BEC configuration. We augment these results by  computing the dynamic structure factor in the density and pseudo-spin sector, and discuss its interesting features
that can be experimentally measured through Bragg spectroscopy of such ultra cold-condensate. 
\end{abstract}

\maketitle

\newpage

\section{Introduction}

The low energy excitation spectrum of an ultra cold atomic Bose-Einstein condensate (BEC) \cite{Stringari, Pethick}reveals a plethora of information about the collective behaviour of such macroscopic quantum systems.
For example, the dispersion relation indicate the breakdown of superfluidity when an object moves through such Bose-Einstein condensate (BEC) \cite{bdsup_theory,bdsup_exp1,bdsup_exp2}, the nature of the quasiparticle modes plays a vital role in characterizing critical behavior near the phase transition \cite{Subir}, and most importantly it registers the response of such superfluid to external fluctuations \cite{Roth} within the frame-work of perturbation theory.
Realisation of artificial light-induced spin-orbit coupling \cite{YJlin,Spielman09,Lin09} in such Bose-Einstein condensate (BEC) and 
has appended a new powerful tool to the simulation toolbox \cite{Dalibardcoll} with such ultra cold atomic systems. Naturally, an extensive insight into the novel properties of such systems  can be gained by looking into the behaviour of their collective excitations \cite{Hall_dipole,Hall_Zhai,Rashba_collmodes,RashbaZhai_CM,sumrules_SOC,CM_Hydrodynamic,dipole_exp}. 

Several studies on the effect of collective modes on various properties of such SOC bosonic superfluid in different dimensions had been carried out focussing on certain specific aspects. They include the study of mean-field dynamics of such SOC  BEC in terms of their collective modes in one and two dimension\cite{RashbaZhai_CM,Rashba_collmodes}, the role of dipole oscillation on the collective behaviour of the system as it goes through a phase transition \cite{sumrules_SOC, dipole_exp}, the change in the velocity of sound as a quasi-one dimensional SOC condensate goes through a second-order phase transition 
\cite{anisodynamics}, study of  collective  modes under hydrodynamic approximation \cite{anisodynamics, CM_Hydrodynamic}, and the existence of mean-field ground state with exotic topology and its subsequent evolution\cite{Hall_Zhai}. For a SOC-BEC with one-dimensional spin-orbit coupling, the collective excitations have been experimentally observed using Bragg spectroscopy, both in homogenous \cite{Peter} and trapped configurations \cite{rotmaxon}.
These theoretical and experimental studies clearly points the necessity of detailed investigation of the full spectrum of these SOC bosonic system in realistic trapping geometry for other dimensions, particularly in two spatial dimension, and the subsequent calculation of the dynamic structure factor using these excitations  that can be experimentally measured using the Bragg spectroscopy \cite{BS1,BS2,BS3, BS4, BS5}. 
  
Motivated by this, in this work, we investigate the excitation spectrum in a trapped quasi-two-dimensional SOC-BEC for both miscible and immiscible configurations and present an analysis on the nature of various quasiparticle excitation modes exploring a wide range of excitation spectrum. We evaluated the excitation spectrum 
both as a function of linear momentum as well as angular momentum.We observe  a minima-like dip at finite wave vector in the excitation spectrum and observe smudged amplitudes of the quasiparticles near this minima in the immiscible case. As we plot the spectrum as a function of angular momentum, 
we see that the the first eigen value in the excitation spectrum consists a non-zero value, compared to the other cases, such as the single component (scalar) BEC, and, two component BEC without SOC. Subsequently we study  the consequences of the Bdg spectrum by  computing the dynamic structure factor in the linear-response regime, a quantity that can be explored in experiments using Bragg spectroscopy.  
Accordingly, the outline of the paper is as follows. In section \ref{Model}, we discuss briefly the model system taken under consideration. In section \ref{BdGformalism}, we discuss the Bogoliubov-de-Gennes (BdG) excitation spectrum for this configuration and characterize the various quasiparticle excitations for various range of momenta's and energy's. We also discuss the features of the low-lying excitations such as dipole and quadrupole modes in this system. In section \ref{strufac}, we compute the static and dynamic structure factor  using which one can experimentally probe excitations in the trapped SOC-BEC. 
\section{Model Hamiltonian for SOC BEC and their low energy collective excitations}
In this section we spell out the details of the methodology of computing the excitation spectrum of a SOC-BEC. In the first part sec. \ref{Model}, we shall introduce the the nature of spin-orbit coupling, the model Hamiltonian, and the corresponding Gross-Piatevskii equation. In sec. \ref{BdGformalism} 
we introduce the details of the Bogoliubov de Gennes formalism applied to such SOC BEC. 
\subsection{Synthetic spin-orbit coupling  for an ultra cold atom and the Gross-Piteavskii equation}\label{Model} 
We will consider the spin-orbit coupled single-particle Hamiltonian, possessing a non-abelian gauge potential of the form $\bs{A}$ : $\it{m}(\eta \check{\sigma}_y,\eta'\check{\sigma}_z,0$) \cite{brandon}, given by: 
\bea
\hat{h}&=&\frac{\bs{\hat{p}}^2}{2 m}\check{I}-\eta \hat{p}_x \check{\sigma}_y-\eta' \hat{p}_y \check{\sigma}_z
\eea
 with $\bs{\hat{p}}=\{\hat{p}_{x},\hat{p}_{y},\hat{p}_{z}\}$; $\check{\sigma}$ and $ \check{I}$'s are Pauli and Identity matrices respectively and $\eta$,$\eta'$ are the strengths of spin-orbit coupling (SOC). The interaction effects are incorporated through the two-body mean field interaction term $1/2\sum_{\kappa\kappa'}\int d{\bf r} V^{\kappa\kappa'}_{int} n_\kappa n_{\kappa'}$ where `$n_\kappa$' is the density of the `$\kappa$'-component, $V^{\kappa\kappa'}_{int}=4\pi \hbar^2 a_{\kappa\kappa'}/m$ correspond to the coupling constants between different spin channels and $a_{\kappa\kappa'}$ represents the $s$-wave scattering length.
 
 The full Hamiltonian within the framework of Gross-Pitaevskii (GP) theory, after projecting the single-particle Hamiltonian into the lower energy subspace, satisfies the following spinorial time-dependent Gross-Pitaevskii equation (see \cite{SOC_SBH} for details):
\bea 
 i\hbar\frac{\partial \psi_{\kappa}  }{\partial t} & = & \bigg[\frac{\hbar^2}{2m}(-i\partial_x- \kappa\frac{m \eta}{\hbar})^2-\frac{\hbar^2}{2m_y}\partial_y^2 + V_{Trap}\nn \\
 & \mbox{+} &  \frac{g+g_{+-}}{2}\bm{\psi^\dagger}\bm{\psi}+\frac{g-g_{+-}}{2}(\bm{\psi^\dagger}\check{\sigma}_z\bm{\psi }) \kappa\bigg]\psi_{\kappa}  \label{gpe} 
\eea
where $\bm{\psi}=[\psi_{+}\hspace{0.2cm}
\psi_{-}]^T$ and $\kappa=\pm$ labels the two components. Also, $m_y = \frac{m}{(1-(\frac{\eta'}{\eta})^2)}$ is the effective mass along the y-
direction (with $\eta'<\eta$) and, $V_{Trap}=m (\omega_x^2 x^2 +\omega_y^2 y^2)/2$ is the external trapping potential with trapping frequencies $\omega_{x,y}$ along x and y directions respectively. Here, we have considered tight confinement along the z-direction and thus, frozen the dynamics along the z-axis \cite{Quasi2D}. The harmonic trap parameters considered in our simulations
are $\omega_x=\omega_y=2\pi \times $ 4.5 Hz ($=\omega_\rho$), $\omega_z=2\pi \times $ 123 Hz. The inter and intraspecies interaction strengths for this quasi-two-dimensional condensate are denoted by $g_{+-}$ and $g_{\kappa\kappa}$ respectively;  $g_{\kappa\kappa'}= \frac{V^{\kappa\kappa'}_{int}}{\sqrt{2\pi}a_{\perp}}$ where $a_\perp = \sqrt{\frac{\hbar}{m\omega_z}}$ is the transverse harmonic oscillator length. 

In Eq.(\ref{gpe}), we have considered the intra- species interaction strengths to be equal $g_{++} = g_{--} (=g$), which is a good approximation to the experimental situation with $^{87}$Rb atoms \cite{g_equal} chosen in this work.
At various places in the paper, we compare our results to a single-component BEC \cite{Stringari,Pethick,DM1} and a two-component BEC \cite{Stringari,Pethick,Myatt,Gordan, Angom}  as limiting cases of Eq.(\ref{gpe}).
It corresponds to substituting the parameters $g_{+-}=0$, $m_y=m$, and $\eta=0$ in Eq.(\ref{gpe}), for the single-component BEC. For a two-component BEC  we set in Eq.(\ref{gpe}), $g_{+-}\neq0$, $m_y=m$, and $\eta=0$.
Additionally we consider the miscible, i.e., 
$g_{+-}<g$ and immiscible configurations, i.e., $g_{+-}>g$ \cite{SOC_Angela,SOC_XIAO}, for a given strength of the spin-orbit coupling. The condensate components overlap each other in the miscible configuration, whereas they are spatially-separated in the immiscible configuration. 

\subsection{Bogoliubov de-Gennes formalism}\label{BdGformalism}
We numerically propagate the GP Eq.($\ref{gpe}$) in imaginary time to obtain the two-component ground state wavefunction of the condensate, $\bm{\psi_0}(x,y)=[\psi_{+,0}(x,y)\hspace{0.2cm}
\psi_{-,0}(x,y)]^T $. The ground state solution obtained from the GPE in this way, for various cases, are shown in Fig. \ref{SFD1} and Fig. \ref{SFD2}.  
The excitation spectrum and the nature of the quasiparticle amplitudes in such SOC-BEC, are then obtained in the framework of the Bogoliubov theory \cite{Stringari,Pethick} by considering the fluctuations over the ground state wavefunction,   of the GP Eq.(\ref{gpe}) as:
\begin{equation}
\resizebox{1\hsize}{!}{$
\begin{bmatrix}
\psi_{+}\\
\psi_{-}\\
 \end{bmatrix} = e^{\frac{-i\mu t}{\hbar}}\bigg(\begin{bmatrix}
\psi_{+,0}\\
\psi_{-,0}\\
 \end{bmatrix}
+ \sum_j \bigg\{\begin{bmatrix}
u_{+,j}\\
u_{-,j}\\
 \end{bmatrix} e^{-i\omega_j t} + \begin{bmatrix}
v^*_{+,j}\\
v^*_{-,j}\\
 \end{bmatrix} e^{i\omega^*_j t}\bigg\}\bigg)  \label{p1}$}
\end{equation}
 where $\mu$ is the chemical potential of the condensate and, $\omega_j$ = $\epsilon_j/\hbar$ where $\epsilon_j$ is the energy corresponding to each quasiparticle excitation. 
 
\begin{figure*}
\includegraphics[scale=0.625]{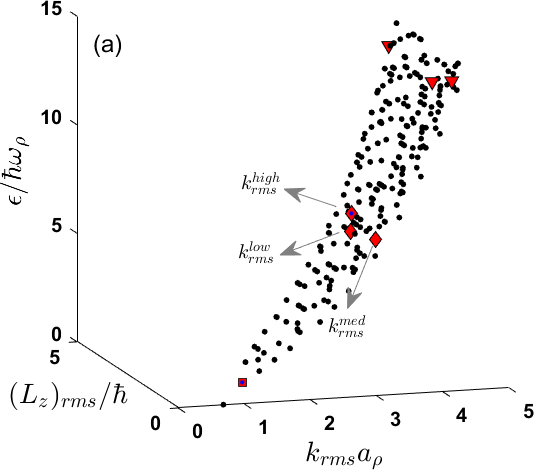}
\includegraphics[scale=0.625]{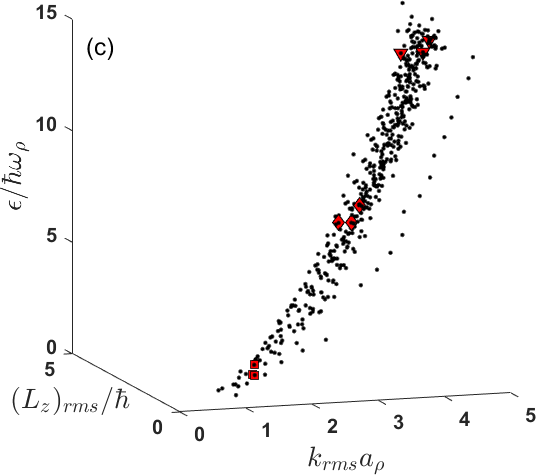}\\
\includegraphics[scale=0.555]{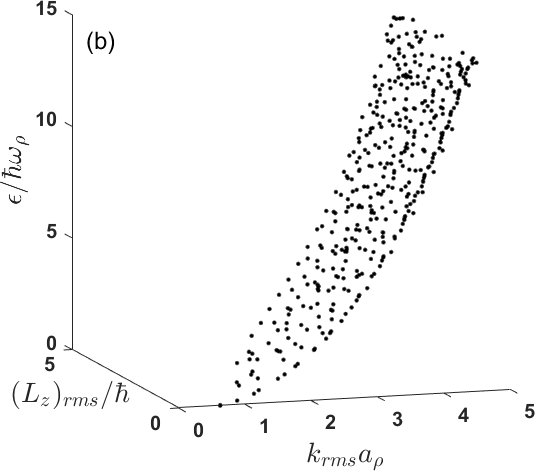}
\includegraphics[scale=0.58]{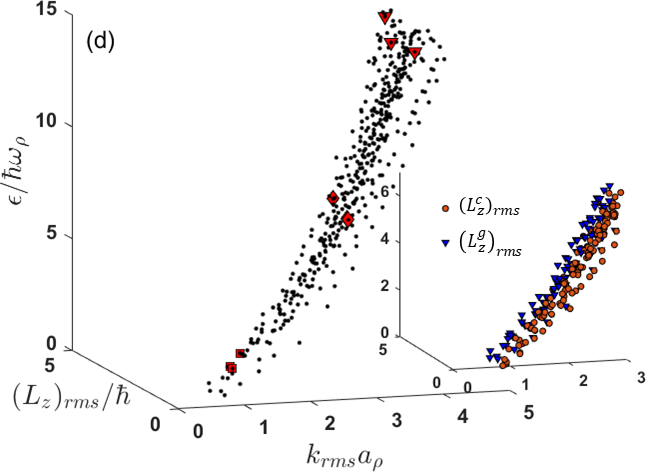}
\caption{{\it (color online).} Dispersion relation as a function of both $k_{rms}$ and ($\hat{L}_z )_{rms}$ : (a) shows the single-component BEC case with $g_{+-} = 0$, (b) shows the dispersion relation for two-component BEC with $g_{+-} = g$. (c-d) shows the dispersion relation for $\eta'/\eta = 0.4$ and $0.78$ respectively, with $\eta = 0.6 \sqrt{\hbar\omega_\rho/m}$ and $g_{+-} = g$. Inset of (d) shows the canonical and gauge contribution to the $rms$ value of total angular momentum.  
}\label{EKm}
\end{figure*}
  The index `$j$' represents the sequence of the quasiparticle
excitation. `$u_{\kappa,j}$'
and `$v_{\kappa,j}$' are spatially dependent complex functions denoting the Bogoliubov
quasiparticle amplitudes corresponding to the $j$th energy eigenstate and are normalized as,
\bea
\int \int dx dy \sum_{\kappa=\pm} [|u_{\kappa,j}(x,y)|^2-|v_{\kappa,j}(x,y)|^2] &=& 1
\eea
Inserting $\psi_\pm({\bf r},t)$ using Eq.(\ref{p1}) into GP Eq. (\ref{gpe}) and retaining the fluctuations upto the linear order, we get the Bogoliubov-de-Genes (BdG) equations for the considered SOC-BEC system:
\begin{widetext}
\begin{eqnarray}
\begin{bmatrix}
\bar{ L}_+&  g \psi_{+,0}^2& g_{+-} {\psi^*_{-,0}} {\psi_{+,0}} & g_{+-} {\psi_{-,0}} {\psi_{+,0}}\\
-g {\psi^*_{+,0}}^2 &\bar{\bar{L}}_+ &-g_{+-}\psi^*_{-,0} \psi^*_{+,0} &-g_{+-}{\psi_{-,0}} \psi^*_{+,0}\\
g_{+-} \psi^*_{+,0} \psi_{-,0}&g_{+-} \psi_{-,0} \psi_{+,0}&\bar{ L}_-&g {\psi^2_{-,0}}\\
-g_{+-} \psi^*_{-,0} \psi^*_{+,0}&-g_{+-}\psi_{+,0} \psi^*_{-,0}&-g{\psi^*_{-,0}}^2&\bar{\bar{L}}_-\\
\end{bmatrix} \begin{bmatrix}
u_{+,j}\\
v_{+,j}\\
u_{-,j}\\
v_{-,j}\\
\end{bmatrix} = \epsilon_j \begin{bmatrix}
u_{+,j}\\
v_{+,j}\\
u_{-,j}\\
v_{-,j}\\
 \end{bmatrix}\label{BdGsolution}
\end{eqnarray}
\end{widetext}
where,
\bea
\bar{ L}_+ &=& \bigg(\frac{\hbar^2}{2m}[-\partial_x^2+ \frac{m^2 \eta^2}{\hbar^2}+2 i \frac{m\eta}{\hbar} \partial_x]-\frac{\hbar^2}{2m_y}\partial_y^2 + V_{Trap}\bigg)\nn\\
&+& (2 g |\psi_{+,0}|^2 + g_{+-}|\psi_{-,0}|^2) - \mu_+ \nn\\
\bar{\bar{L}}_+ &=& -\bar{ L}_+\nn\\
\bar{ L}_- &=& \bigg(\frac{\hbar^2}{2m}[-\partial_x^2+ \frac{m^2 \eta^2}{\hbar^2}-2 i \frac{m\eta}{\hbar} \partial_x]-\frac{\hbar^2}{2m_y}\partial_y^2 + V_{Trap}\bigg)\nn\\
&+& (2 g |\psi_{-,0}|^2 + g_{+-} |\psi_{+,0}|^2) - \mu_- \nn\\
\bar{\bar{L}}_- &=& -\bar{ L}_-\nn
\eea
We diagonalise the above BdG matrix \ref{BdGsolution} numerically to find quasiparticle excitation energy $\epsilon_j$,
as well as the Bogoliubov quasiparticle amplitudes \{$u_{ij}$ and $v_{ij}$\}.
To this end we expand $u_{ij}$ and $v_{ij}$ in terms of harmonic oscillator eigenfunction $\phi_n(x), \phi_n(y) $  for such harmonically trapped SOC-BEC (details given in Appendix \ref{integral}) as 
\bea
u_{+,j}(x,y)= \sum_{k,l=0}^{N_b} p_{jkl} \phi_{k}(x)\phi_{l}(y)\nn\\
v_{+,j}(x,y)= \sum_{k,l=0}^{N_b} q_{jkl} \phi_{k}(x)\phi_{l}(y)\nn\\
u_{-,j}(x,y)= \sum_{k,l=0}^{N_b} r_{jkl} \phi_{k}(x)\phi_{l}(y)\nn\\
v_{-,j}(x,y)= \sum_{k,l=0}^{N_b} s_{jkl} \phi_{k}(x)\phi_{l}(y)\label{qa}
\eea
where $p_{jkl}$, $q_{jkl}$, $r_{jkl}$ and $s_{jkl}$ are the coefficients of the linear combination, and,
 $\phi_n(x) = \frac{1}{\sqrt{2^n {n!}a_\rho \sqrt{\pi}}} e^{-x^2/2a_\rho^2} H_n(x/a_\rho)
$; where $H_n$ is the $n$th order Hermite polynomial and $a_\rho = \sqrt{\frac{\hbar}{m\omega_\rho}}$. Substituting the above expression, after some straightforward algebra, the resulting equation can be written as

\bea
\begin{bmatrix}
[A] &  [B] & [C]  & [D] \\
[E] &  [F] & [G]  & [H] \\
[I] &  [J] & [K]  & [L] \\
[M] &  [N] & [O]  & [P] \\
\end{bmatrix} = \epsilon \begin{bmatrix}
[P]\\
[Q]\\
[R]\\
[S]\\
 \end{bmatrix}\label{BDG}
\eea
where, the each matrix [.] is of the dimension $(N_b+1)^2\times (N_b+1)^2$ and the elements of each of these matrices are evaluated using the integrals defined in Appendix \ref{integral}. The resulting BdG matrix in Eq.(\ref{BDG}) has $4(N_b + 1) \times 4(N_b + 1)$ dimensions for equal number of basis (=$N_b$) along x and y directions. The size of the BdG matrix increases rapidly due the $\sim N_b^2$ scaling. We obtain the converged eigenvalues (upto four digits after decimal) for $N_b$ = 50.

To construct the dispersion relation of the Bogoliubov
modes, we adopt the methodology used for the trapped configurations such as two-component BEC  \cite{Ticknor} and dipolar condensates \cite{Wilson, Blakie1,Blakie2,Blakie3}. 
Since there is no translational symmetry in such finite, trapped configuration, the linear momentum is not a good quantum number. Thus each quasiparticle can be linked to an $rms$ value of the linear momentum. Additionally 
the irrotationality condition that is obeyed in single component BEC-superfluid with no SOC coupling gets violated in  SOC coupled two-component BEC superfluid \cite{diffvort}. However, the angular momentum operator $\hat{L}_{z}$ also does not commutes with the Hamiltonian in Eq.(\ref{gpe}). 
So, to see the consequence we evaluate the $rms$ value of the z-component of angular momentum
for each such excitation \cite{ang}. 
Accordingly we define
\bea
k^j_{rms} &=& \bigg(\frac{\sum_\kappa \int d{\bf k} {\bf k}^2 [|u_{\kappa, j}({\bf k})|^2 +|v_{\kappa, j}({\bf k})|^2] }{\sum_\kappa \int d{\bf k} [|u_{\kappa, j}({\bf k})|^2 +|v_{\kappa, j}({\bf k})|^2]}\bigg)^{1/2}\label{krms}
\eea
\begin{equation}
\resizebox{1.05\hsize}{!}{($\hat{L}^j_z )_{rms}= \bigg(\frac{\sum_\kappa \int d{\bf r} [u^*_{\kappa, j}({\bf r})\hat{L}^2_z u_{\kappa, j}({\bf r}) +v^*_{\kappa, j}({\bf r})\hat{L}_z^2 v_{\kappa, j}({\bf r})]) }{\sum_\kappa \int d{\bf r} [|u_{\kappa, j}({\bf r})|^2 +|v_{\kappa, j}({\bf r})|^2]}\bigg)^{1/2}$}\label{lz}
\end{equation}
where $\kappa= +,-$ labels the two components.  Here, in Eq.(\ref{krms}),
$u_{\kappa, j}({\bf k}) = \mathcal{F} [u_{\kappa, j}({x,y})]$ and $v_{\kappa, j}({\bf k}) = \mathcal{F} [v_{\kappa, j}({x,y})]$ are the quasiparticle amplitudes in the momentum space. Also, the angular momentum operator can be written as $\hat{L}_z = \hat{L}^c_z +\hat{L}^g_z $. The effective momentum along a particular direction (along x) in Eq.(\ref{gpe}) gets modified due to spin-orbit coupling; thus, the angular momentum operator also contains additional terms apart from the canonical one. Here, $\hat{L}^c_z = (x\hat{p}_y - y \hat{p}_x)$ corresponds to the canonical part, and $\hat{L}^g_z = \kappa$($\eta y m$) represents the spin-dependent gauge part.
 \begin{figure}
\includegraphics[scale=0.75]{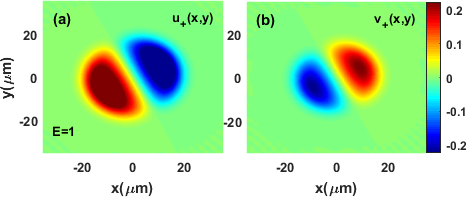}
\includegraphics[scale=0.75]{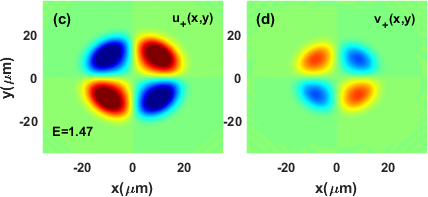}
\caption{{\it (color online).} Quasi-particle amplitudes $u_+$ and $v_+$, for the dipole (a,b) and quadrupole modes (c,d) is shown for the single-component BEC with intra-species interaction strength $g_{+-}=0$. $u_+$ and $v_+$ are in units of $a_\rho^{-1}$.
}\label{DMQM_1}
\end{figure}
\begin{figure}
\includegraphics[scale=0.7]{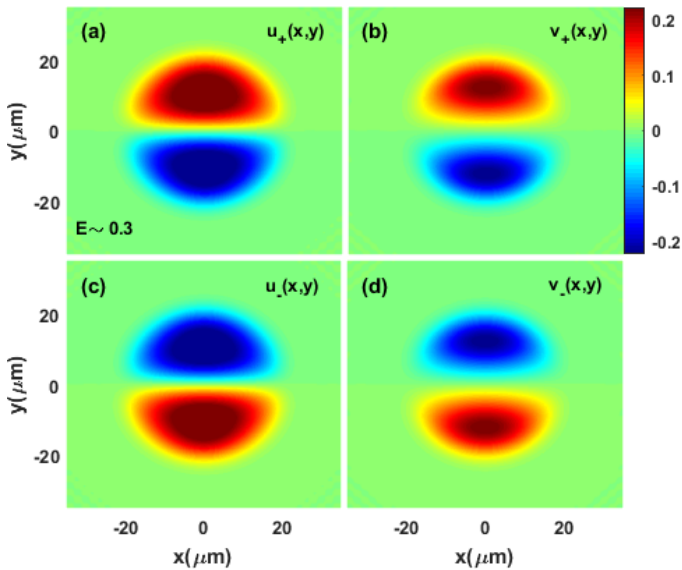}
\caption{{\it (color online).} Quasi-particle amplitudes $u_\pm$ and $v_\pm$, for the dipole mode is shown for the case with intra-species interaction strength $g_{+-}\neq 0$ and $\eta'/\eta = 0.78$. $u_+$ and $v_+$ are in units of $a_\rho^{-1}$.
}\label{DM2}
\end{figure}

The excitation spectrum as a function of $rms$ value of both the linear and angular momentum is shown in Fig. \ref{EKm}. Fig. \ref{EKm} (a) illustrate the excitation spectrum for the single-component case, with intra-species  interaction strength $g_{+-} = 0$, and Fig. \ref{EKm} (b) shows the spectrum for the two-component BEC with no SO coupling and $g_{+-} = g$. Fig. \ref{EKm} (c-d) shows the excitation spectrum in two-component SOC-BEC for lower and higher ratio of spin-orbit coupling strengths $\eta'/\eta$ =  0.4 and 0.78 respectively.

\begin{figure}
\includegraphics[scale=0.6]{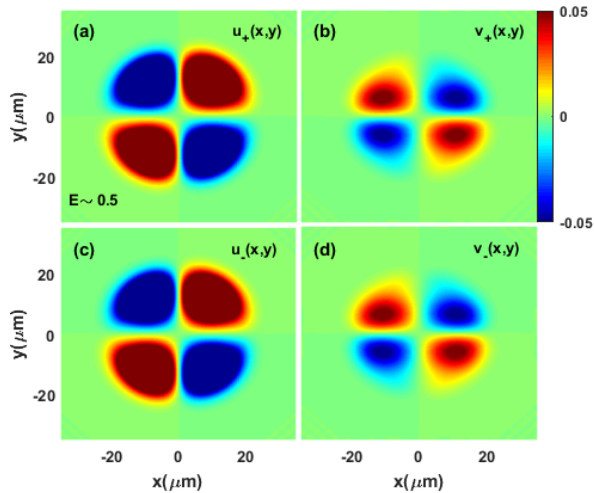}
\caption{{\it (color online).} Quasi-particle amplitudes $u_\pm$ and $v_\pm$, for the quadrupole mode is shown for the case with intra-species interaction strength $g_{+-}= g$ and $\eta'/\eta = 0.78$. $u_+$ and $v_+$ are in units of $a_\rho^{-1}$.
}\label{QM2}
\end{figure}

We start our discussion with the low-lying dipole and quadrupole excitations. Collective dipole oscillation represents the center-of-mass motion of all the atoms and are the excitations corresponding to the lowest finite energy modes \cite{DM1}.
In order to excite these modes, the trap can be suddenly displaced, in such a way that the initial ground state of the BEC is now
no longer the ground state of the displaced trap. This results to collective excitation of the atoms and thus, can be easily excited experimentally \cite{DM_exp}. For the first case shown in Fig. \ref{EKm} (a), the dipole mode occurs at energy $\epsilon = \hbar \omega_\rho$. This is also in consonance with the fact that for a scalar condensate, the frequency of the dipole oscillation is just the harmonic-trap
frequency \cite{DM1,DM_rmp}. However, for a SO-coupled condensate, the dipole-oscillation frequency deviates
from the trap frequency \cite{RashbaZhai_CM,dipole_exp,sumrules_SOC,diffvort}, which is also consistent with our simulations. For example, the energy at which the dipole mode occur in SOC-BEC for strengths $\eta'/\eta=0.78$ is $E=\epsilon/\hbar\omega_\rho = 0.3$, which is different from 1.

We next investigate the structure of the quasiparticle amplitudes to interpret the physics behind the dispersion curve. For comparison, the quasiparticle amplitudes \{$u_+,v_+$\} of the dipole and quadrupole mode for the case of single-component BEC are showed in Fig. \ref{DMQM_1} (a,b) and (c,d), respectively.  Examining the dipole mode and quadrupole mode behavior provides a signature of the quantum phase transition and is studied in SOC-BEC experimentally \cite{DPT}. We here show the quasiparticle amplitudes for both components labeled through $\pm$, \{$u_\pm,v_\pm$\} for dipole and quadrupole modes, for one of the SOC strengths $\eta'/\eta=0.78$ with non-zero intra-species interaction strength, in Fig. \ref{DM2} (a-d) and Fig. \ref{QM2} (a-d), respectively. In both these figures, (a-b) shows the quasiparticle amplitudes for the `+' component, and (c-d) contains the quasiparticle amplitudes for the `-' component. 

\begin{figure}
\hspace*{-1.5cm}\includegraphics[scale=0.57]{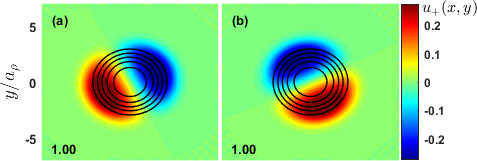}\\
\includegraphics[scale=0.57]{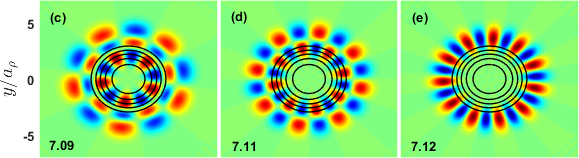}\\
\includegraphics[scale=0.57]{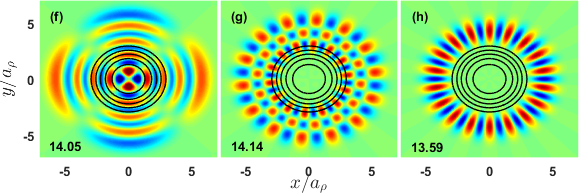}\\
\caption{{\it (color online).} Quasi-particle amplitudes $u_+(x,y)$ (in units of $a_\rho^{-1}$) in the uncoupled scalar condensate (without SO coupling) with $g_{+-} = 0$, for few excitations marked in Fig. \ref{EKm} (a), is shown. Here (a-b), (c-e) and (f-h) shows the quasiparticle amplitude corresponding to the excitations marked with \protect\markerfour, \protect\markerfive \hspace{0.1cm}and \protect\markerthree \hspace{0.1cm}respectively. In each set shown here, quasiparticles begin with small values of $k_{rms}$ (at left) and move
towards large values of $k_{rms}$ (at right). The value of dimensionless energy $\epsilon/(\hbar \omega_\rho)$ is written at the left corner (at bottom) in each figure. Each sub-figure is superimposed with the condensate density contours (black lines) corresponding to the $`+'$ component.
}\label{modes_nonsoc}
\end{figure}

After discussing the low energy excitations, we will now examine the nature of the excitations for higher energy values. For illustration, we evaluate the quasiparticle amplitudes at dimensionless energies $\epsilon/(\hbar \omega_\rho)$ near to 1 (low), 7 (intermediate) and 14 (high), marked with \protect\markerfour, \protect\markerfive \hspace{0.1cm}and \protect\markerthree \hspace{0.1cm}respectively. The three energy ranges are chosen to examine the whole range of energies in the dispersion (Fig. \ref{EKm}) exhibited by the cases under consideration. 

For each energy range chosen, we now discuss the effect of different $k_{rms}$ values on the quasiparticle amplitudes.  We choose a particular excitation, at a given energy,  each from the rightmost and leftmost point, from the dispersion curve in Fig. \ref{EKm}, and one in between these two. The leftmost excitation (with smaller $k_{rms}$) that lies nearly on the linear part of the dispersion, is phonon-like. The rightmost one (with larger $k_{rms}$) corresponds to the surface excitation \cite{DM1}, as we shall see the quasiparticle amplitude gets distributed over the boundary of the condensate only for such modes.  

Again for comparison, we discuss the case of a scalar condensate, whose quasiparticle amplitudes ($u_+$) are displayed, in Fig. \ref{modes_nonsoc}. We will analyze the radial and angular nodes in various examples of quasiparticle excitations. The radial nodes are the nodes existing in each quasiparticle excitation mode along any cross-section from the origin (${\bf r} = 0$), i.e., nodes along $r = \sqrt{(x^2+y^2)}$ counts $(n_r)$ except the zero at large $r$. The number of azimuthal nodes ($n_\phi$) is equal to twice the maxima (or minima) number along the outermost boundary of the excitation amplitude containing both. Fig. \ref{modes_nonsoc}(a-b) shows two quasiparticles with energy, $\epsilon/(\hbar \omega_\rho)$ = 1. These degenerate modes are marked, in Fig. \ref{EKm} (a) with red squares. Both of these have one radial node and two angular nodes. 
 
In Fig. \ref{modes_nonsoc} (c-e), we show quasiparticles with energy, $\epsilon/(\hbar \omega_\rho)$ $\sim$ 7 for three illustrative values of $k_{rms}$: low (c), intermediate (d) and high (e) value. The excitation at lower value of $k_{rms}$ lies nearly on the linear part of the dispersion curve, has three radial nodes and 10 angular nodes. The intermediate $k_{rms}$ value at the same value of energy has two radial nodes and 16 angular nodes, shown in Fig. \ref{modes_nonsoc} (d). Whereas, larger $k_{rms}$ value at the same value of energy has only one radial node and 22 angular nodes. Therefore, roughly at same range of energy, the quasiparticle amplitude gains azimuthal nodes at the cost of loss in radial nodes. This is also evident from Fig. \ref{EKm}   (a), where the quasiparticle excitation mode with smaller $k_{rms}$ has lower value of ($\hat{L}_z )_{rms}$ and the quasiparticle excitation mode with larger $k_{rms}$ value has higher ($\hat{L}_z )_{rms}$ value. The observation holds for other considered cases also. 

Additionally, Fig. \ref{EKm} illustrates that for a scalar BEC and a two-component BEC, the lowest quasiparticle excitation has zero $rms$ value of angular momentum. 
 However, for the SOC case, the gauge part of the $rms$ value of angular momentum ($L_z^g$) results in contributing non-zero angular momentum to the lowest-energy excitation mode, shown in Fig. \ref{EKm} (d) (inset) for one of the SOC strengths. It occurs due to the modification of the effective momentum along a particular direction [Eq. (\ref{gpe})], subjected to variation in spin-orbit coupling parameters, which results in an anisotropic nature of the SOC-BEC's velocity profile. In turn, this leads to violation of irrotationality condition on the velocity fields in the SOC- BEC superfluid, which is in contrast to the case of scalar BEC superfluid \cite{diffvort}. Hence, the lowest energy excitation in SOC-BEC, as a consequence, have a finite $rms$ value of angular momentum. 
\begin{figure}
\hspace*{-0.15cm}\includegraphics[scale=0.55]{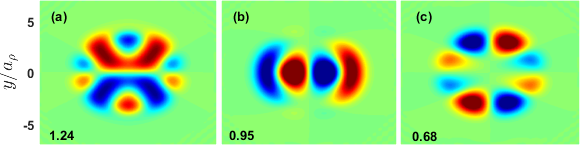}\\
\hspace*{-0.15cm}\includegraphics[scale=0.565]{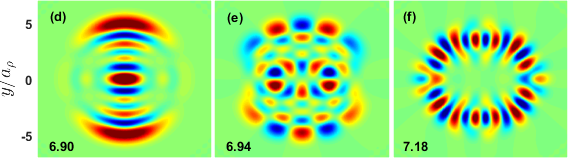}\\
\includegraphics[scale=0.565]{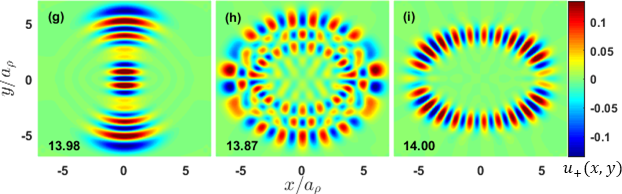}\\
\caption{{\it (color online).} Quasiparticle amplitudes $u_+(x,y)$ (in units of $a_\rho^{-1}$) with moderate ratio of SO coupling strengths $\eta'/\eta = 0.4$ with $\eta = 0.6\sqrt{\hbar\omega_\rho/m}$, for few excitations marked in Fig. \ref{EKm} (c), is shown. Here (a-c), (d-f) and (g-i) shows the quasiparticle amplitude corresponding to the excitations marked with \protect\markerfour, \protect\markerfive \hspace{0.1cm}and \protect\markerthree \hspace{0.1cm}respectively. In each set shown here, quasiparticles begin with small values of $k_{rms}$ (at left) and move
towards large values of $k_{rms}$ (at right). The value of dimensionless energy $\epsilon/(\hbar \omega_\rho)$ is written at the left corner (at bottom) in each figure.
}\label{modes_lowrsoc}
\end{figure}
\begin{figure}
\hspace{150cm}\includegraphics[scale=0.555]{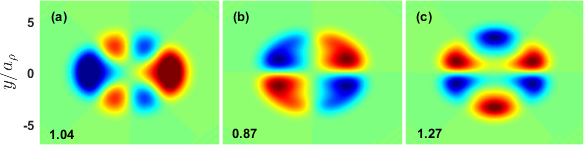}\\
\includegraphics[scale=0.56]{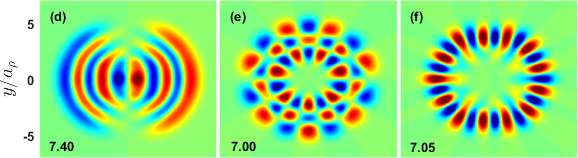}\\
\includegraphics[scale=0.558]{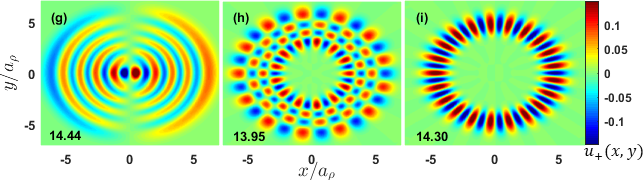}\\
\caption{{\it (color online).} Quasi-particle amplitudes $u_+(x,y)$ (in units of $a_\rho^{-1}$) with SO coupling strengths ratio $\eta'/\eta = 0.78$ with $\eta = 0.6\sqrt{\hbar\omega_\rho/m}$, for few excitations marked in Fig. \ref{EKm} (d), is shown. Here (a-c), (d-f) and (g-i) shows the quasiparticle amplitude corresponding to the excitations marked with \protect\markerfour, \protect\markerfive \hspace{0.1cm}and \protect\markerthree \hspace{0.1cm}respectively . In each set shown here, quasiparticles begin with small values of $k_{rms}$ (at left) and move
towards large values of $k_{rms}$ (at right). The value of dimensionless energy $\epsilon/(\hbar \omega_\rho)$ is written at the left corner (at bottom) in each figure.
}\label{modes_highrsoc}
\end{figure}
\begin{figure}
\includegraphics[scale=0.56]{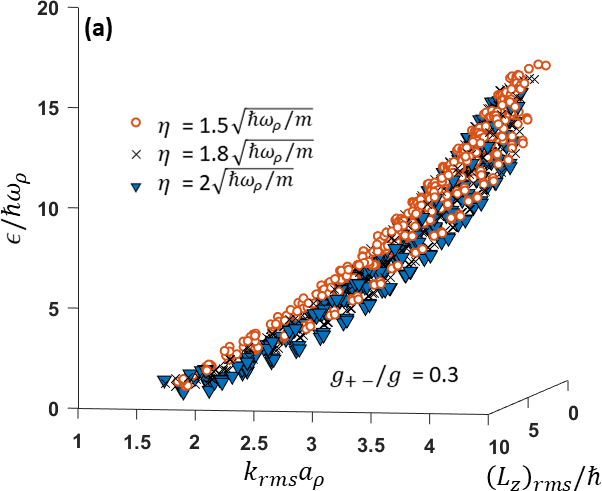}\\
\vspace{0.5cm}
\includegraphics[scale=0.56]{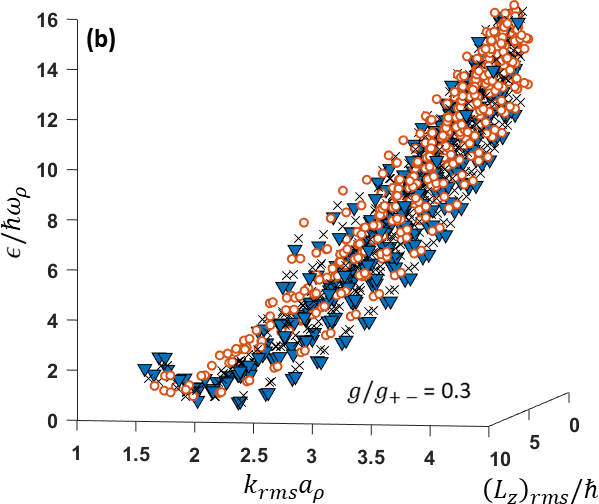}\\
\vspace{0.5cm}
\includegraphics[height=4.4cm, width=9cm]{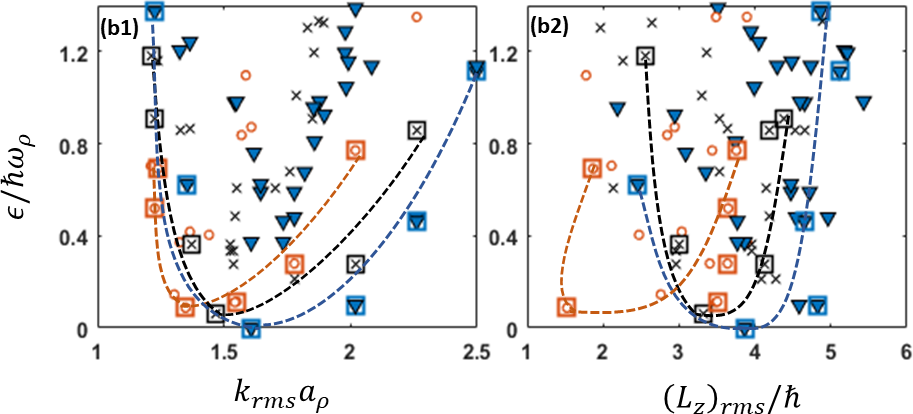}
\caption{{\it (color online).}  Excitation spectrum for  various SOC strengths $\eta = 1.5, 1.8, 2$ (in units of $\sqrt{\hbar\omega_\rho/m}$) with $\eta'/\eta = 0.78$, in the miscible (a) and immiscible (b) cases, are shown. (b1)-(b2) respectively shows the magnified part of the dispersion near the minima, in miscible configuration, concerning $k_{rms}$ and $(L_z)_{rms}$ separately. Some points on the outermost branch of the dispersion, viewed w.r.t $k_{rms}$, are marked (using squares). The corresponding points are marked in (b2) also, and dashed lines guide the eye joining these.}\label{EK_numerical}
\end{figure}

Also, at the same range of energy value, but with increasing value of $k_{rms}$ the quasiparticle amplitudes move towards the outermost boundary of the condensate and away from the center of the trap. To illustrate this we have superposed the quasiparticle amplitude for a particular component `$u_+(x,y)$' with the contour lines of the corresponding condensate density i.e, for the `+'-component $|\psi_+|^2$. It is shown in Fig. \ref{modes_nonsoc} (c-h).  At relatively larger values of $k_{rms}$ shown in Fig. \ref{modes_nonsoc} (e,h), the excitation amplitude lies at the outermost edge/contour of the condensate and thus, corresponds to the surface mode. It shows the coexistence of surface and phonon-like modes is observed at the similar range of energy. The observation is also consistent for the energy value $\sim$ 14 at low, intermediate and high value of $k_{rms}$. (f) has 8 radial nodes and 4 angular nodes; (g) has 3 radial nodes and 28 angular nodes; and (h) has only a single radial node but 36 angular nodes. 

\begin{figure*}
\hspace*{-0.4cm}\includegraphics[scale=0.63]{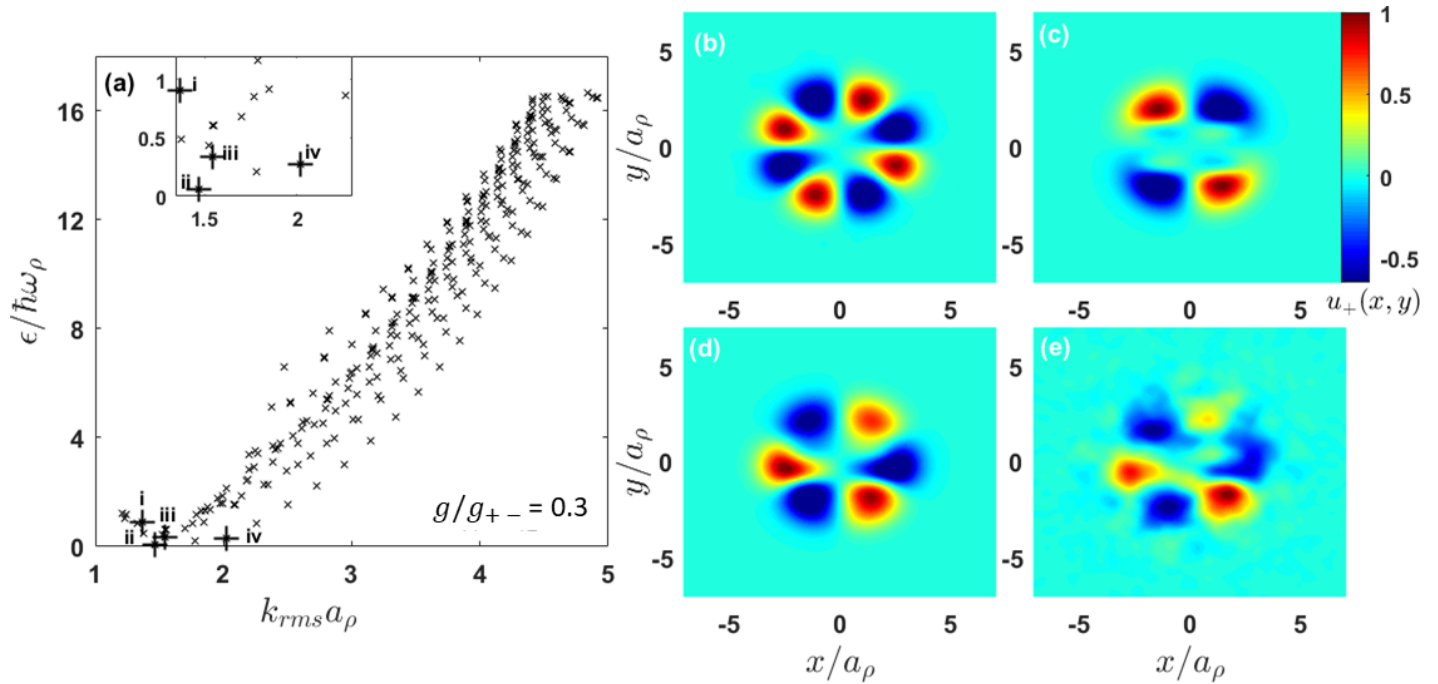}
\caption{{\it (color online).} {\it Quasiparticle amplitudes for modes near the minima}: (a) displays the excitation spectrum for miscible case from Fig. \ref{EK_numerical} (b), for ratio of SOC strengths $\eta'/\eta = 0.78$ with $\eta = 1.8$ (in units of $\sqrt{\hbar\omega_\rho/m}$). Inset shows the magnified view near the minima. (b-e) respectively shows the quasiparticle amplitude $u_+(x,y)$ for a few modes in (a), labelled as (i-iv) with marker `+'.
}\label{Qampgamma18e1}
\end{figure*}

After discussing the nature of the quasiparticle amplitudes of the various excitations in a scalar condensate, we next analyze the nature of these in our SO coupled configuration. Fig. \ref{modes_lowrsoc}-\ref{modes_highrsoc} shows the quasiparticle amplitude $u_+(x,y)$ at different energy values for SOC strengths $\eta'/\eta = 0.4, 0.78$, respectively. Table \ref{table1} contains the details about the number of radial nodes and angular nodes in the quasiparticle amplitude for the cases considered. Some points of differences and similarities between SOC and scalar condensate, based on Table \ref{table1}, are discussed in the following. Firstly, in a SOC-BEC, at low $k_{rms}$ and considered energy values, the quasiparticle excitations are aligned along a specific direction. It is along x-direction in the case $\eta'/\eta = 0.78$, shown in Fig.\ref{modes_highrsoc}(d,g). It is along y-direction in the case $\eta'/\eta = 0.4$, shown in Fig. \ref{modes_lowrsoc}(d,g). Whereas, for the case of scalar condensate, the distribution of amplitude is uniform and symmetric, shown in Fig. \ref{modes_nonsoc}(c,f). The anisotropy in quasiparticle amplitudes in SOC-BEC is present, although the trap is isotropic. It occurs due to the different effective masses and the effective momentum along x and y directions in  Eq. (\ref{gpe}). Also, in a SOC-BEC, the number of nodes is direction-dependent and is not constant. As an example, consider the case illustrated in Fig. \ref{modes_lowrsoc}(e) for one of the SOC-strengths ratios at $\epsilon/\hbar\omega_\rho = 7$. The number of radial nodes at intermediate $k_{rms}$ consists of either 3 or 4 radial nodes depending on the direction chosen.  

Furthermore, at high $k_{rms}$, the number of angular nodes for $\epsilon/\hbar \omega_\rho = 7,14$  increases in the order: single component, $\eta'/\eta = 0.78$, $\eta'/\eta = 0.4$. At the same value of $\epsilon/\hbar \omega_\rho = 7,14$ but at low $k_{rms}$, the number of angular nodes is only 2 for both $\eta'/\eta = 0.78, 0.4$ and is largest for single-component BEC. Also, the number of radial nodes for $\epsilon/\hbar \omega_\rho = 1,7,14$, at high $k_{rms}$, is mostly 1 for all the cases. The number of radial nodes for $\epsilon/\hbar \omega_\rho = 7,14$, at low $k_{rms}$, is greater in SOC cases than the single-component BEC.

\begin{table}
\begin{center}
 \begin{tabular}{| c | c | c | c |} 
 \hline
 $n_r$ & Scalar & $\eta'/\eta = 0.4$ & $\eta'/\eta = 0.78$\\ [0.5ex] 
 \hline\hline
 $\epsilon/\hbar\omega_\rho$ $\sim$ 1 & 1 & (1/2,2,1) & (1,1,1) \\ 
 \hline
  $\epsilon/\hbar\omega_\rho$ $\sim$ 7 & (3,2,1) & (6,3/4,1/2) & (5,3,1) \\
 \hline
   $\epsilon/\hbar\omega_\rho$ $\sim$ 14 & (8,3,1) &   (9,3/4,1) & (9,4,1)\\[1ex] 
 \hline
\end{tabular}
\end{center}
\begin{center}
 \begin{tabular}{| c | c | c | c |} 
 \hline
 $n_\phi$ & Scalar & $\eta'/\eta = 0.4$ & $\eta'/\eta = 0.78$\\ [0.5ex] 
 \hline\hline
  $\epsilon/\hbar\omega_\rho$ $\sim$ 1 & 2 & (10,2,8) & (6,4,6) \\ 
 \hline
  $\epsilon/\hbar\omega_\rho$ $\sim$ 7 & (10,16,22) & (2,10,30) & (2,14,26) \\
 \hline
   $\epsilon/\hbar\omega_\rho$ $\sim$ 14 & (4,28,36) & (2,30,46) & (2,26,42) \\[1ex] 
 \hline
\end{tabular}
\caption{\label{table1} The number of radial and angular nodes $n_r$ and $n_\phi$ respectively are listed in order ($k^{low}_{rms},k^{med}_{rms},k^{high}_{rms}$) for various cases.}
\end{center}
\end{table}

Fig. \ref{EK_numerical}(a-b) shows the dispersion relation in SOC-BEC for various SOC strengths $\eta = 1.5, 1.8, 2$ (in units of $\sqrt{\hbar\omega_\rho/m}$) with $\eta'/\eta = 0.78$ in miscible, i.e., 
$g_{+-}<g$ and immiscible configurations, i.e., $g_{+-}>g$ respectively. 

The excitation spectrum in the immiscible system [Fig. \ref{EK_numerical}(b)] contains a minima-like feature at finite $k_{rms}$. A few outermost points (branch) of the dispersion, viewed against $k_{rms}$, are marked with squares. The minima's location gets right-shifted towards a higher value of $k_{rms}$ with an increase in the magnitude of $\eta$, shown in Fig. \ref{EK_numerical}(b1). We also show the magnified part of the dispersion from Fig. \ref{EK_numerical}(b) concerning $(L_z)_{rms}$ in Fig. \ref{EK_numerical}(b2).  The points corresponding to (b1) are marked in (b2) also. Fig. \ref{EK_numerical}(b2) also contains minima at finite $rms$ value of angular momentum, demonstrating that the lowest energy excitation have a non-zero, finite angular momentum, that increases with an increase in $\eta$'s magnitude.

We next investigate the nature of the quasiparticle modes near the minima. To this purpose, we consider the dispersion relation from Fig. \ref{EK_numerical}(b1) concerning $k_{rms}$ only. It is shown, in Fig. \ref{Qampgamma18e1}(a), for the parameters $g<$ $g_{+-}$ and $\eta'/\eta = 0.78$ with $\eta = 1.8\sqrt{\hbar\omega_\rho/m}$. Fig. \ref{Qampgamma18e1}(b-e) shows the quasiparticle amplitude $u_+(x,y)$ for the modes marked with `+' in (a), for four illustrative modes labeled as (i-iv). The mode marked with `(ii)' identifies the minima location and has four angular nodes. Fig. \ref{Qampgamma18e1}(b) with the label (i), has eight angular nodes. Both (d-e) marked with (iii-iv), respectively, have six angular nodes. Thus, the considered  case for investigating quasiparticle modes near the minima has no radial nodes but only angular nodes. Here, the number of the angular nodes is least at the minima location, with the largest on its left. Moreover, the effect of unbalanced interaction strengths, i.e., $g_{+-}\neq g$ also impacts the spread of the quasiparticle amplitudes which become smeared spatially, illustrated in Fig. \ref{Qampgamma18e1}(b-e).

\section{Structure Factor}\label{strufac}
Using the  excitations spectrum of a trapped SOC-BEC computed in the previous section, we can now calculate the dynamic structure
factor  (DSF), that can be measured using a robust experimental tool, Bragg spectroscopy \cite{BS1,BS2,BS3, BS4, BS5}. DSF yields the
information on the spectrum of collective excitations, which can be explored at low momentum transfer \cite{BS3}. It also provides knowledge about
the momentum distribution through which the behavior
of the system at high momentum transfer can be characterized, where the response is dominated by single-particle effects \cite{BS1}.
Its investigation has provided a crucial understanding of physics in superfluid ${}^4$He \cite{Griffin}. Specially, it has facilitated the measurement of the roton spectrum\cite{Hespectrum}, pair-distribution function and condensate fraction available from neutron scattering experiments \cite{Griffin2}, in this system. 
\begin{figure}
\hspace*{-2.55cm}\includegraphics[scale=0.5]{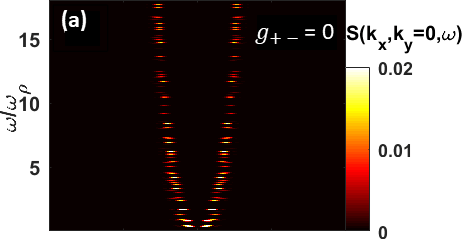}\\\vspace{0.2cm}
\includegraphics[scale=0.42]{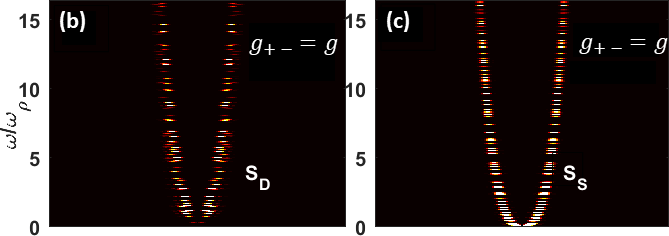}\\\vspace{0.2cm}
\includegraphics[scale=0.5]{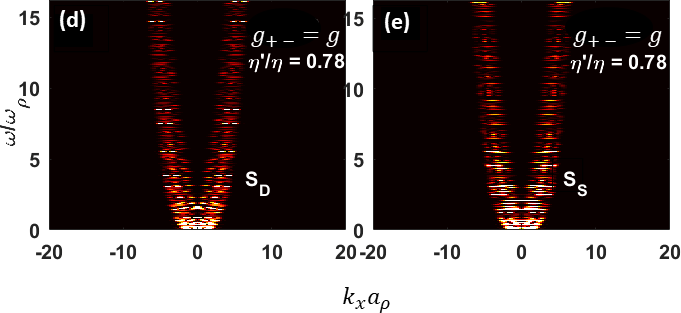}
\caption{{\it (color online).} (a) shows the dynamic structure factor for single-component BEC with respect to $k_x (= k)$ and $\omega$. (b-c) shows the dynamic structure factor for density $S_D(k_x,k_y = 0, \omega)$ and spin $S_S(k_x,k_y = 0, \omega)$ in two-component BEC. (d-e) shows these for SOC case for strengths $\eta'/\eta = 0.78$ with $\eta = 0.6\sqrt{\hbar\omega_\rho/m}$.}\label{SF}
\end{figure}

\begin{figure}
\hspace*{-0.60cm}\includegraphics[scale=0.42]{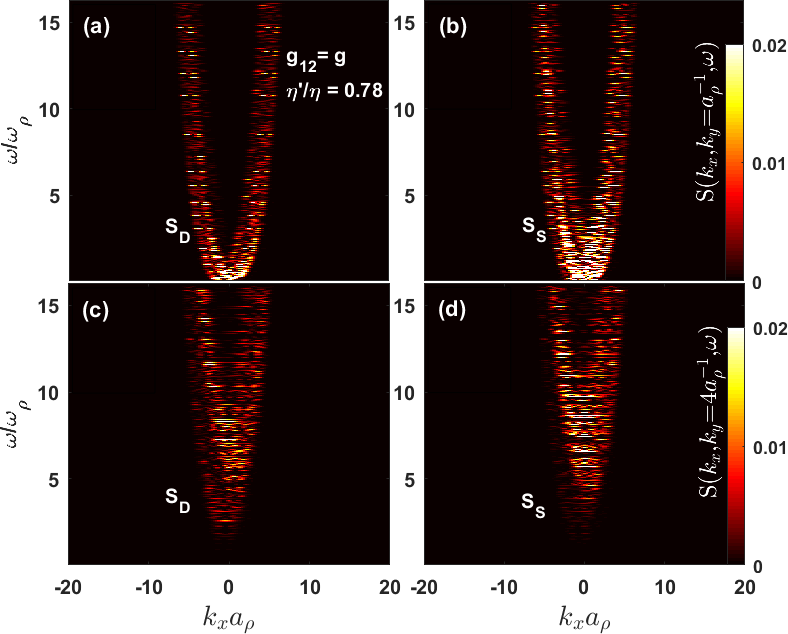}
\caption{{\it (color online).} (a) shows the dynamic structure factor for single-component BEC with respect to $k_x (= k)$ and $\omega$. (b-c) shows the dynamic structure factor for density $S_D(k_x,k_y = 0, \omega)$ and spin $S_S(k_x,k_y = 0, \omega)$ in two-component BEC. (d-e) shows these for SOC case for strengths $\eta'/\eta = 0.78$ with $\eta = 0.6\sqrt{\hbar\omega_\rho/m}$. 
}\label{SF}
\end{figure}

\begin{figure}
\hspace*{-1.55cm}\includegraphics[scale=0.52]{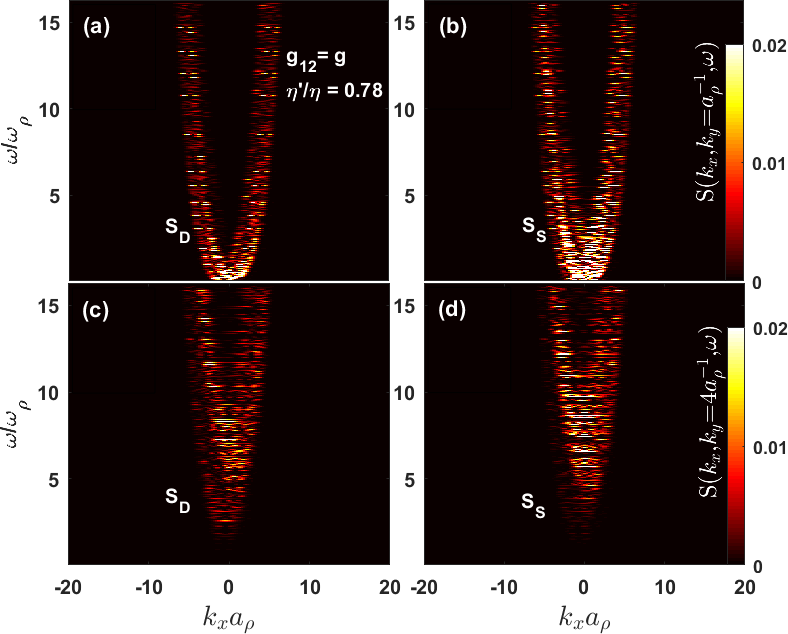}
\caption{{\it (color online).} (a,b) shows the density and spin dynamic structure factor in a SOC-BEC for strengths $\eta'/\eta = 0.78$ with $\eta = 0.6\sqrt{\hbar\omega_\rho/m}$ at $k_y=a_\rho^{-1}$. The corresponding results at $k_y =4a_\rho^{-1}$ are shown (c,d) respectively. 
}\label{DSF_vary_ky}
\end{figure}

\begin{figure}
\centering
\includegraphics[scale=0.7]{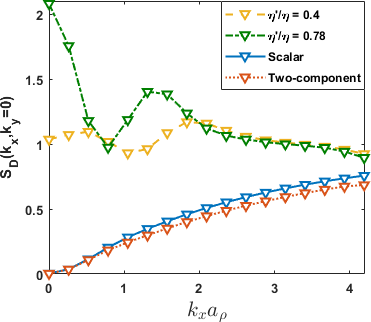}
\includegraphics[scale=0.7]{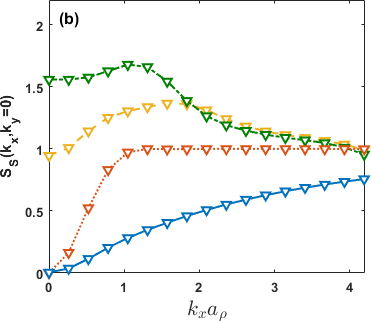}
\caption{{\it (color online).} (a-b) respectively shows the density $S_D(k_x, k_y=0)$ and spin static structure factor $S_S(k_x, k_y=0)$ for various listed cases. 
}\label{SSSF}
\end{figure}

\begin{figure}
\centering
\includegraphics[scale=0.660]{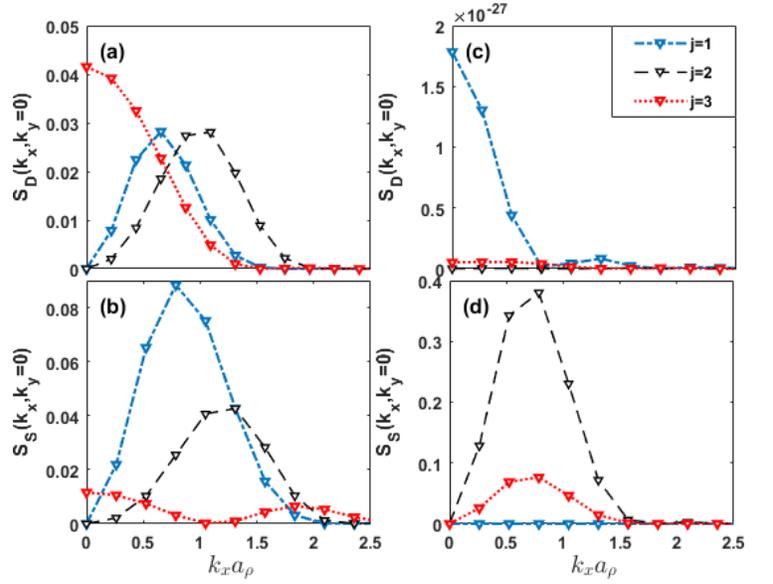}
\caption{{\it (color online).} (a-b) shows contribution of the first three excited states (j=1,2,3) to the total density and spin static structure factor in SOC-BEC, respectively, for $\eta'/\eta=0.78$; (c-d) shows the same for the two-component BEC.
}\label{SSSFc}
\end{figure}
\begin{figure}
\includegraphics[scale=0.812]{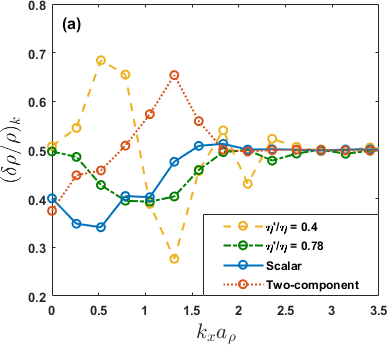}\\
\includegraphics[scale=0.812]
{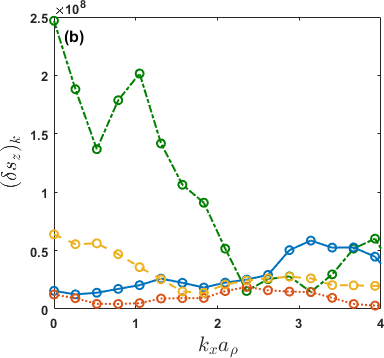}\caption{{\it (color online):} (a) displays the Fourier transform of the fluctuations in the total density relative to the total density, for several cases mentioned. (b) shows the Fourier transform of the spin-density fluctuations.
}\label{Fluctuation}
\end{figure}

In the experiment, the condensate is  impinged by a Bragg pulse, with the help of two laser beams having wavevectors ${\bf k_1}$ and ${\bf k_2}$ and a frequency difference $\omega$. The following Hamiltonian describes the resulting external perturbation term:
\bea
\hat{H}_{pert}(t) = \lambda \hat{G}_{\bf k}^\dagger e^{-i\omega t}e^{\gamma t} +H.c.\label{BP}
\eea
where $\lambda$ is the strength of the Bragg potential, ${\bf k }= {\bf k_1}-{\bf k_2}$ is the momentum transferred by the Bragg pulse to the condensate and the factor $e^{\gamma t}$ in Eq.(\ref{BP}), with $\gamma \rightarrow 0^{+}$, ensures that the system at initial time ($t=-\infty$) is governed by the unperturbed Hamiltonian.
Directly after perturbing the system, the DSF relative to the operator $\hat{G}_{\bf k}$ \cite{Stringari} can be probed:
\bea
S({\bf k},\omega) &=& \sum_j Z_j \delta(\hbar\omega-\hbar\omega_{j0})\label{DSF1}
\eea
 The quantity $Z_j = |\bra{j}\hat{G}_{\bf k}\ket{0}|^2$ is known as the strength of the operator $\hat{G}_{\bf k}$ with respect to state $\ket{j}$ and $\hbar \omega_{j0} = \epsilon_j-\epsilon_0$. Here $\ket{0}$ ($\ket{j}$) correspond to the ground (excited) state with the energy
$\epsilon_0$ ($\epsilon_j$). For the two-component SOC-BEC, the DSF corresponding to the total density $\hat{\rho}({\bf r}) = \hat{\bm{\psi}} ^\dagger({\bf r}) \hat{\bm{\psi}}({\bf r})$ and spin density $\hat{s}_z ({\bf r}) = \hat{\bm{\psi}} ^\dagger({\bf r}) \hat{\sigma}_z\hat{\bm{\psi}}({\bf r})$ can be also calculated from the BdG excitations computed in previous section, where $\hat{\bm{\psi}}({\bf r})$ is defined in Eq.(\ref{foperator}). Here, the operator $\hat{G}_{\bf k}$, corresponding to the total density is $\hat{\rho}_{\bf k} = \int d{\bf r}\hat{\bm{\psi}} ^\dagger ({\bf r})\hat{\bm{\psi}}({\bf r})e^{i{\bf k}\cdot{\bf r}}$ and, the operator corresponding to the spin density is $\hat{s}_{z_{\bf k}} = \int d{\bf r} \hat{\bm{\psi}} ^\dagger({\bf r}) \hat{\sigma}_z\hat{\bm{\psi}}({\bf r})e^{i{\bf k}\cdot{\bf r}}$, where $\hat{\bm{\psi}}(\bf r)=[\hat{\psi}_{+}(\bf r)\hspace{0.2cm}
\hat{\psi}_{-}(\bf r)]^T $. 

The strength's $Z_j$ in Eq.(\ref{DSF1}) can be simplified and written in terms of the qausiparticle amplitudes (details in Appendix \ref{appC2}), resulting DSF corresponding to total density (spin density), labelled with subscript `$_D$'(`$_S$'), given as:
\begin{equation}
\resizebox{1.05\hsize}{!}{$
S_{D}(k_x,k_y,\omega) = \sum_j \big{|}\int dx dy e^{i(k_xx+k_yy)} \bra{0} \hat{\rho}({\bf r}) \ket{j}\big{|}^2  \delta(\hbar\omega - \hbar\omega_j)$}\nn
\end{equation}
\begin{equation}
\resizebox{1.05\hsize}{!}{$
S_{S}(k_x,k_y,\omega) = \sum_j \big{|}\int dx dy e^{i(k_xx+k_yy)} \bra{0} \hat{s}_z({\bf r}) \ket{j}\big{|}^2  \delta(\hbar\omega - \hbar\omega_j)\label{dsf}$}
\end{equation}
The  DSF integrated over  $S(k_x,k_y,\omega)$ over frequency domain,  gives the static structure factor,  
\beq S(k_x,k_y) = \frac{1}{N}\int d\omega S(k_x,k_y,\omega) \nonumber \eeq

We start by discussing the analytical results for the structure factor in a uniform scalar and two-component BEC. The dispersions in these two cases are respectively given by 
\bea \epsilon_B & = & \hbar\sqrt{c^2k^2+ (\hbar k^2/2m)^2} \label{single}  \\
\epsilon_B^{D,S} & = & \hbar\sqrt{c_{D,S}^2k^2+ (\hbar k^2/2m)^2} \label{twocomp} \eea 
In a uniform scalar BEC, the sum in Eq.(\ref{DSF1}) is expended by only one mode. The static structure factor in this system is $S({\bf k}) = \epsilon_0({\bf k})/\epsilon_B({\bf k})$ ($Feynman$ relation) \cite{Feynman}, where $\epsilon_0({\bf k})$ is the dispersion of non-interacting bosons. In the limit $k \rightarrow 0$, $S({\bf k})$ = $\hbar k/ 2mc$ where $c=\sqrt{gn/m}$ is the Bogoliubov sound velocity.
 This can be generalised in a uniform two-component BEC using the dispersion relation  \cite{Timmermann} given in Eq. (\ref{twocomp}).  The static factor in such configuration for $k \rightarrow 0$ gives $S_{D,S}({\bf k})$ = $\hbar k/ 2mc_{D,S}$, where $c_{D,S}$ represents the sound velocity corresponding to density and spin-density, respectively. Thus, as $k \rightarrow 0$, the static structure factors in both  configurations vary linearly with wavevector. However, the static structure factor approaches unity at large momentum \cite{BS2,BS3} in both of the systems, which is the value of the static structure factor for any momentum in an uncorrelated 
non-interacting atoms. 



To benchmark our computation for the trapped SOC-BEC, we now evaluate the dynamic and the static structure factor in the trapped single and two-component BEC by considering the suitable limit of  the excitation energy and the quasiparticle amplitudes determined in section \ref{BdGformalism}. Substituting parameters $g_{+-}=0$, $m_y=m$, and $\eta=0$ in Eq.(\ref{gpe}) we get the results for trapped scalar BEC. The dynamic structure factor $S(k_x,k_y = 0,\omega)$ for this case is shown Fig.\ref{SF}(a). To get a clear insight into its behavior, we integrate the dynamic structure factor $S(k_x,k_y,\omega)$ over the frequency domain and compute the static structure factor. The evaluated structure factor is illustrated, with a bold (blue) line in Fig. \ref{SSSF}(a,b) (Here, $S_D =S_S =S$). The substitution of parameters $g_{+-}\neq0$, $m_y=m$, and $\eta=0$ in Eq.(\ref{gpe}) corresponds to the case of a trapped two-component BEC, discussed in \cite{Ticknor2}. In the case of two-component BEC, the density and spin dynamic structure factor are represented, in Fig. \ref{SF} (b-c), respectively. The density static structure factor's general behavior is identical to that of the single-component BEC, illustrated in Fig. \ref{SSSF}(a).

The magnitude of $S_{D}(k_{x}, k_{y})$ in two-component BEC is relatively lower than that of the single-component BEC at most $k_x$ values for the uniform and trapped case. Similarly $S_{S}(k_{x}, k_{y})$ has a higher magnitude relative to the corresponding single-component static structure factor. It reaches the plateau of the unit static structure factor rapidly, shown in Fig. \ref{SSSF}(b). Thus, finite interspecies interaction strength diminishes the density static structure factor in two-component BEC, whereas it intensifies the spin structure factor, in agreement with results of \cite{Ticknor2}. Again this can be understood from $S({\bf k}) = \epsilon_0({\bf k})/\epsilon_B({\bf k})$. For a finite $g_{+-}$, the effective interaction strength ($g+g_{+-}$) increases in the total density part [refer to the first term in the second line of Eq.(\ref{gpe})], leading to increment in the energy of the excitation relative to the single-component case with $g_{+-}=0$ and thus decreasing the static structure factor. Simultaneously, the effective interaction strength ($g-g_{+-}$) is reduced in the spin-density term [refer to the second term in the second line of Eq.(\ref{gpe})], resulting in the lower energy and increasing the static structure factor relative to $S_D$ in two-component BEC. 

However, the $Feynman$ relation discussed above is not generally applicable to uniform BEC's with spin-orbit coupling \cite{FR_SOCBEC}. The relation is not satisfied in the whole momentum space, and applicability depends on the type of ground state of SOC-BEC \cite{FR_SOCBEC2}. Therefore, we will utilize Eq. (\ref{dsf}) only to compute and discuss the dynamic structure factor features in trapped SOC-BEC.
 \begin{figure*}
\includegraphics[scale=0.8]{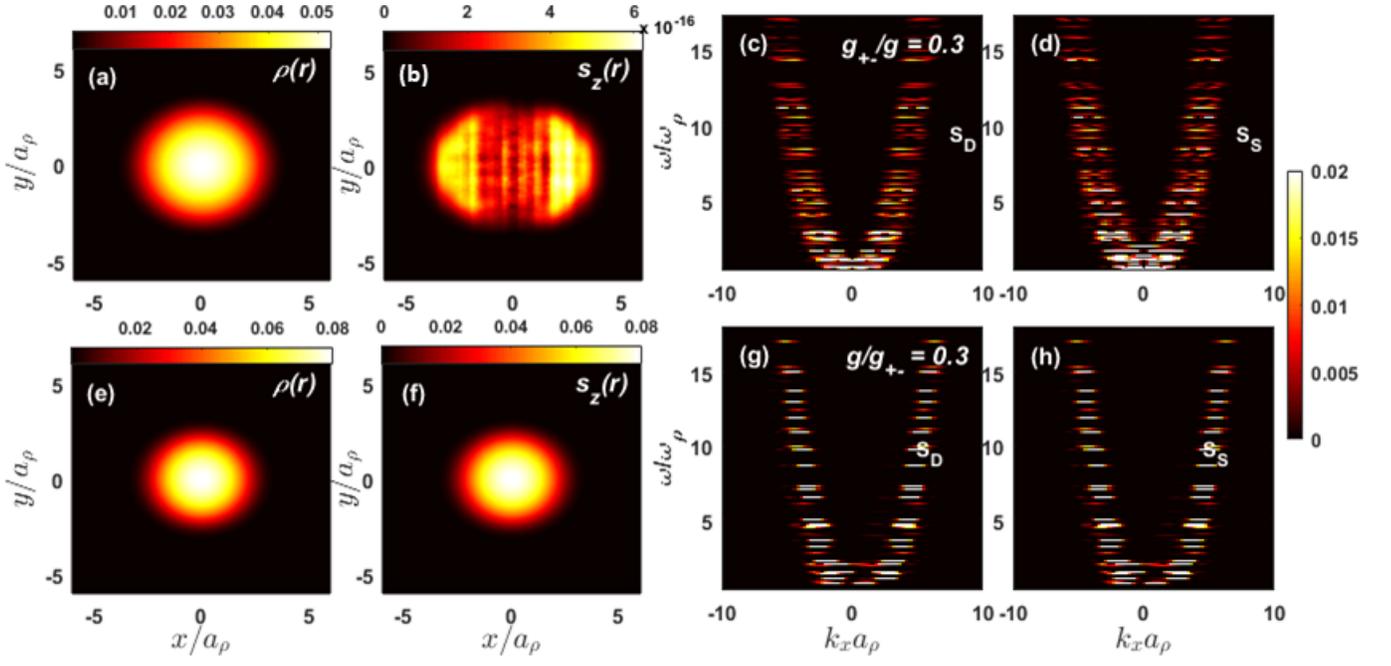}
\caption{{\it (color online).} (a-b) respectively shows the ground state corresponding to the total density and spin-density (in units of $a_\rho^{-2}$) in miscible SOC configuration ($g_{+-} < g$) for $\eta'/\eta = 0.78$ with $\eta = 0.6\sqrt{\hbar\omega_\rho/m}$. The dynamic structure factor for density $S_D(k_x,k_y = 0, \omega)$ and spin $S_S(k_x,k_y = 0, \omega)$ in this case is shown in (c-d) respectively. (e-h) shows the corresponding results for $g_{+-} > g$.
}\label{SFD1}
\end{figure*}
\begin{figure*}
\includegraphics[scale=0.82]{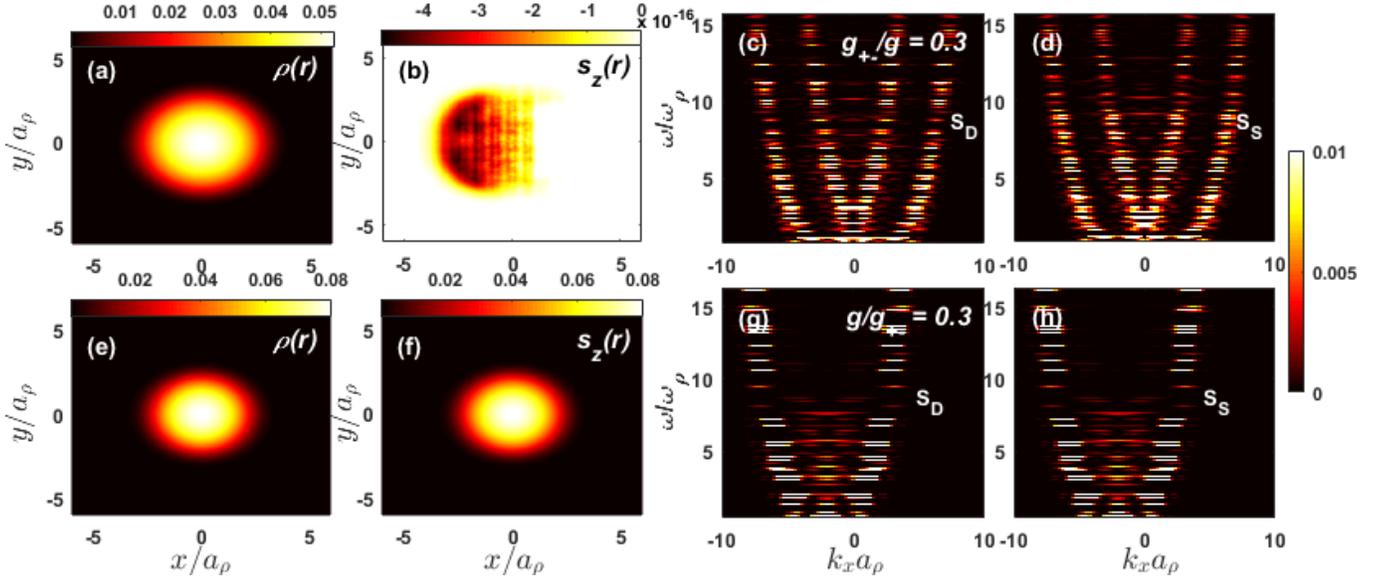}
\caption{{\it (color online).} (a-b) respectively shows the ground state corresponding to the total density and spin-density (in units of $a_\rho^{-2}$) in miscible SOC configuration ($g_{+-} < g$) for $\eta'/\eta = 0.78$ with $\eta = 2\sqrt{\hbar\omega_\rho/m}$. The dynamic structure factor for density $S_D(k_x,k_y = 0, \omega)$ and spin $S_S(k_x,k_y = 0, \omega)$ for this case is shown in (c-d) respectively. (e-h) shows the corresponding results for $g_{+-} > g$.
}\label{SFD2}
\end{figure*}

Coming to the SOC-BEC with all interaction strengths equal, namely $g_{+-} = g$, the $S_{D,S}(k_{x}, k_{y}=0, \omega)$  is shown for $\eta'/\eta = 0.78$ with $\eta = 0.6 \sqrt{\hbar\omega_\rho/m}$ in Fig. \ref{SF}(d-e), respectively. The figures illustrate a relatively broader non-vanishing frequency response at a given wavevector and vice-versa, compared to the case without SOC. The corresponding static structure factors are shown in Fig. \ref{SSSF} (a-b), respectively. In contrast to the single-component and two-component BEC, both the static structure factors are finite in the limit of $k_x\rightarrow$ 0.
For fixed $k_y=a_\rho^{-1},4a_\rho^{-1}$, the dynamic structure factor corresponding to the density and spin dynamic structure factor is shown in Fig. \ref{DSF_vary_ky} (a-b) and (c-d) respectively.

We also illustrate the contribution of the first three excited states, corresponding to j =1,2,3 from Eq.(\ref{DSF1}), 
  individually to the total density and spin structure factor in Fig. \ref{SSSFc}(a-b), respectively. And, compare them with the same for a two-component BEC in Fig. \ref{SSSFc}(c-d). In SOC-BEC, a finite contribution develops in both the static factors $j=3$ onwards, as $k_x \rightarrow 0$. However, in two-component BEC, these terms' contribution to $S_{D,S}(k_{x}, k_{y})$ as $k_x \rightarrow 0$ is negligible, illustrated in Fig. \ref{SSSFc}(c-d), respectively.

Furthermore, the $S_{D}(k_{x}, k_{y})$ in SOC-BEC consists a peak at $k_x \rightarrow 0$ followed by a minima and a maxima as $k_x$ increases, illustrated in Fig. \ref{SSSF}(a). Also, for $S_{S}(k_{x}, k_{y})$, we observe a peak at finite $k_x$, illustrated in Fig. \ref{SSSF}(b). The maxima in the  $S_{D}(k_{x}, k_{y})$ is due to the behavior of the fluctuations in the ground state wave function. The density and spin fluctuation in terms of quasiparticle amplitudes is respectively, given as, 
\bea 
\delta \rho ({\bf r}) &=& \sum_j [|u_{+,j}+v^*_{+,j}|^2 + |u_{-,j}+v^*_{-,j}|^2]\\
\delta s_z ({\bf r})&=& \sum_j [|u_{+,j}+v^*_{+,j}|^2 - |u_{-,j}+v^*_{-,j}|^2]
\eea
 The density static structure factor's maxima coincides approximately with the location of the maximum in the fluctuations in the density relative to the total density, in the momentum space, i.e., $\delta_\rho(k_x,k_y=0)/\rho(k_x,k_y=0)$, illustrated in Fig. \ref{Fluctuation} (a) corresponding to the total density. Fig. \ref{Fluctuation} (b), shows the Fourier transform of the fluctuations, $(\delta s_z)_k$ in the ground state corresponding to the spin-density alongwith $k_y = 0$. It does  not have clear maxima/minima like the previous one [Fig. \ref{Fluctuation} (a)]. However, $S_S$ shown in Fig. \ref{SSSF} (b) for $\eta'/\eta=0.78$ has a distinct peak whose location coincides with the peak in Fig.\ref{Fluctuation} (b) for the corresponding case.

The spin-orbit coupling further enhances the structure factor's amplitude for both the density and spin static structure factor, shown in Fig. \ref{SSSF} (a-b), respectively. Also, the peak's amplitude at non-zero $k_x$ is relatively large in the $S_{S}(k_{x}, k_{y})$ than in $S_{D}(k_{x}, k_{y})$. Similar to the case without SOC, the effective interaction strength gets reduced in the spin-density term which further leads to lower excitation energy and enhancing the spin static structure factor compared to the density structure factor.

The above discussion confines to the balanced inter-component and intra-component interaction strengths, i.e., $g_{\pm} = g$. Based on the interaction strengths, the characterisation of the ground state in an interacting gas depends on the minima in the single-particle dispersion \cite{brandon,SOC_SBH}, which is either a single-well  (miscible, $g< g_{\pm}$  ) or double-well (immiscible, $g > g_{\pm}$)  type.
The ground state of our system has two possible phases: zero-momentum phase and plane-wave phase \cite{Li,Rev}. In the miscible regime, SOC-BEC condenses in the zero-momentum state and has a extremely small value ($\sim$ 0) of spin-density, apparent from the magnitude of randomly fluctuating spin-density, shown in Fig. \ref{SFD1}-\ref{SFD2}(b). Whereas in the immiscible regime, SOC-BEC chooses one of the two-wells as the ground state and is in the plane-wave phase. In this regime, the spin-density is finite-valued \cite{Rev}, illustrated in Fig. \ref{SFD1}-\ref{SFD2}(f). Another typical ground state phase in SOC-BEC literature is the stripe phase \cite{Li}, where the BEC remains in a superposition state of both well. However, the stripe phase is not achievable in our system because there is no off-diagonal coupling term in Eq. (\ref{gpe}).  

We next discuss the density and spin dynamic structure factor and the corresponding static structure factors in SOC-BEC, in such  miscible and 
and immiscible configuration. For both these configurations, we have shown total density and spin-density in the ground state with the respective dynamic factors in Fig. \ref{SFD1}-\ref{SFD2}, for $\eta'/\eta = 0.78$ with $\eta = 0.6,1.8$ (units of $\sqrt{\hbar\omega_\rho/m}$), respectively. 

In the case of $g_{+-} < g$, the density and spin dynamic structure factor $S_{D,S}(k_x,k_y=0,\omega)$ for SOC strengths $\eta'/\eta = 0.78$ with $\eta = 0.6\sqrt{\hbar\omega_\rho/m}$ is shown in Fig. \ref{SFD1} (c-d) respectively, and for $g_{+-} > g$ in Fig. \ref{SFD1} (g-h) respectively. For a higher value of SOC strength $\eta = 2\sqrt{\hbar\omega_\rho/m}$ with $\eta'/\eta = 0.78$, the corresponding results appear in Fig. \ref{SFD2}. At a given SOC-strength and $g_{+-} < g$, the DSF for both total density and spin-density are symmetric about $k_x$ = 0, i.e., $S_{D,S}(k_x,k_y=0,\omega)=S_{D,S}(-k_x,k_y=0,\omega)$. In the opposite case, $g_{+-} > g$, the energy spectrum possesses a minima-like feature, as discussed earlier. Also, SOC-BEC chooses either of the two wells as the ground state through the spontaneous symmetry breaking mechanism, which consequently affects the nature of the DSF \cite{anisodynamics}. The symmetry under the exchange of  $k_x$ to $-k_x$ is now not respected. The DSF exhibits asymmetric character about $k_x$ = 0 in both total density and spin-density, i.e., $S_{D,S}(k_x,k_y=0,\omega) \neq S_{D,S}(-k_x,k_y=0,\omega)$, illustrated respectively in Fig. \ref{SFD1}-\ref{SFD2}(g-h).   

\begin{figure}
\includegraphics[scale=0.7]{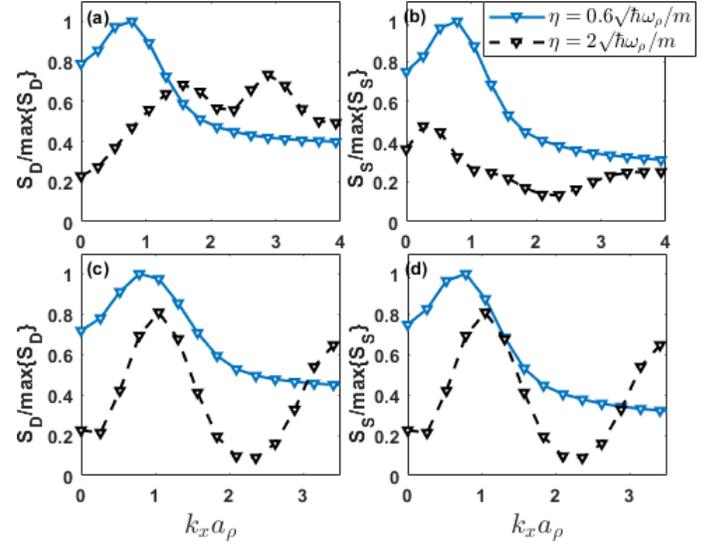}
\caption{{\it (color online).}Static structure factor for density $S_D(k_x,k_y = 0)$ and spin $S_S(k_x,k_y = 0)$ in the SOC-BEC case for ratio SOC strengths $\eta'/\eta = 0.78$ with $\eta = 0.6, 2$ (in units of $\sqrt{\hbar\omega_\rho/m}$), for $g_{+-} < g$ is shown in (a-b) respectively. (c-d) shows the corresponding results for $g_{+-} > g$. Here, the results have been normalized by the maximum value of the corresponding static structure factor.
}\label{SSF}
\end{figure}

Fig. \ref{SSF} displays the static structure factor in both miscible and immiscible SOC-BEC. In the miscible case, for a ratio of SOC strengths $\eta'/\eta = 0.78$ with values of $\eta$ listed in Fig. \ref{SSF}, the density and spin-density static structure factor is shown in (a-b), respectively; (c-d) shows the corresponding results for the immiscible case. The features discussed previously for balanced interaction strengths, e.g., non-zero static structure factor as $k_x\rightarrow 0$ and the magnitude of static structure factor approaching unity at large wave vector, remain valid.

The static structure factors in Fig. \ref{SSF} are normalized by their maximum value, i.e., $S_{D,S}/\text{max}\{S_{D,S}\}$, to show the general features such as peaks, maxima/minima, and not magnitude. The behavior of both the density and spin static structure factors for smaller $\eta$ is similar. It develops a peak at finite $k_x$ and then reaches a constant value. However, the increase in the value of SOC strength $\eta$ results in maxima-minima variation in density and spin static structure factors for $\eta = 2 \sqrt{\hbar\omega_\rho/m}$. Thus, the structure factor's character is strongly affected by the variation in the interaction strengths and the SOC strengths.

\section{Conclusion}
 Within the BdG theory framework, we analyse the excitation spectrum in a trapped quasi-two-dimensional two-component SOC-BEC for both balanced and unbalanced interaction strengths. For comparison, we also computed the excitation spectrum in configurations without spin-orbit coupling, i.e., in the single-component and two-component BEC. Apart from identifying low-energy excitations such as dipole, quadrupole modes in SOC-BEC, and single-component BEC, we examined various excitation modes probing a range of energy spectrum. We particularly demonstrate how the SOC results in the anisotropic quasiparticle amplitudes even in an isotropic trapping potential. We also demonstrated that the gauge part of the total angular momentum is responsible for granting non-zero angular momentum to the lowest-energy excitation in SOC-BEC.  
 
 Furthermore, the immiscible SOC-BEC configuration possesses a minima-like feature in the excitation spectrum. The quasiparticle amplitudes near the minima exhibit smudged-asymmetrical nature due to the imbalanced interaction strengths.  We evaluate the dynamic structure factor to understand the response of all these configurations to the external perturbation, within the BdG framework. Results in section \ref{strufac} display the strong impact of SOC strengths and the interaction strengths on the the dynamic and static structure factor. The results presented can be studied beyond the mean-field theory by taking account of quantum fluctuations, e.g., using the truncated Wigner approximation. Thus, enabling the evaluation of structure factor in SOC-BEC at finite temperature. Moreover, Bragg spectroscopy allows one to tune the momentum transfer over a wide range, and various properties, e.g., coherence \cite{BS1}, and vortices \cite{vort1,vort2} can be probed in such a configuration.

We also show that the lowest energy excitation has a non-zero angular momentum arising from the gauge part. This may lead to the study the Einstein-de Haas effect \cite{EDH, EDH_CM, EDH_BEC} in such system (magnetization present in the initial system can cause mechanical rotation in it). We can tune the SOC parameters that change the non-Abelian gauge potential and hence the effective magnetic field. The magnetic field variation can lead to finite polarisation in the system causing rigid rotation in this system. These may lead to interesting theoretical and experimental work in future.

\section{ACKNOWLEDGMENTS}
 The work is supported by a BRNS (DAE, Govt. of India)
Grant No. 21/07/2015-BRNS/35041 (DAE SRC Outstanding Investigator scheme). 
\appendix
\section{Details of Basis chosen and Integrals used in section \ref{BdGformalism}}\label{integral}
Consider the following part of $\bar{ L}_+$ from Eq. (\ref{BdGsolution}),
\bea
\bigg[\frac{1}{2m} (p_x+m\eta)^2+\frac{1}{2}m\omega^2x^2 \bigg]+\bigg[\frac{1}{2m_y} p_y^2+\frac{1}{2}m\omega^2 y^2 \bigg] \nn
\eea
The above can be rewritten as:
\bea
\bigg[P_1^2+X_1^2 \bigg]+\bigg[P_2^2+X_2^2\bigg] \nn
\eea
where, $P_1=\frac{1}{\sqrt{2m}}(p_x+m\eta)$, $P_2=\frac{p_y}{\sqrt{2m_y}}$, $X_1=\sqrt{\frac{m}{2}}\omega x$ and  $X_2=\sqrt{\frac{m}{2}}\omega y$.
Defining the operators:
\bea
a_1'&=& \frac{1}{\sqrt{\hbar \omega}}(X_1+iP_1), a_2'= \frac{1}{\sqrt{\hbar \omega}}({\frac{m}{m_y}})^{1/4}(X_2+iP_2)\nn
\eea
We get,
\bea
X_1^2 +P_1^2 &=&  \bigg[ a_1'^\dagger  a_1' + \frac{1}{2} \bigg] \hbar \omega,\nn\\
X_2^2 +P_2^2 &=& \bigg[ a_2'^\dagger  a_2' + \frac{1}{2} \bigg] \hbar \omega\sqrt{\frac{m}{m_y}}\nn
\eea
  
Using these operators, first part of $\bar{ L}_+$ (or $\bar{\bar{L}}_+)$ becomes diagonal in the chosen basis with energy $$E(K,L)=(K+ m_R L + \frac{1}{2} [1+m_R] )\hbar \omega$$ where $m_R = \sqrt{m/m_y}$ and K,L =0,1,...$N_b$. 

We next provide the explicit form of the integrals used for computing the matrix elements of Eq.(\ref{BDG}) in the following:
\bea
A_{k'l'kl} &=& \int \int dx dy \phi^*_{k'} \phi^*_{l'}[\bar{L}_+] \phi_{k}\phi_{l}\nn\\
B_{k'l'kl} &=& \int \int dx dy \phi^*_{k'} \phi^*_{l'}[g_{2D} {\psi^2_{+,0}}] \phi_{k}\phi_{l}\nn\\
C_{k'l'kl} &=& \int \int dx dy \phi^*_{k'} \phi^*_{l'}[g_{2D} \psi_{+,0}{\psi^*_{-,0}}] \phi_{k}\phi_{l}\nn\\
D_{k'l'kl} &=& \int \int dx dy \phi^*_{k'} \phi^*_{l'}[g_{2D} \psi_{+,0}\psi_{-,0}]\phi_{k}\phi_{l}\nn\\
E_{k'l'kl} &=& \int \int dx dy \phi^*_{k'} \phi^*_{l'}[-g_{2D} {\psi^*_{+,0}}^2]\phi_{k}\phi_{l}\nn\\
G_{k'l'kl} &=& \int \int dx dy \phi^*_{k'} \phi^*_{l'}[- g_{2D} {\psi^*_{+,0}}{\psi^*_{-,0}}] \phi_{k}\phi_{l}\nn\\
H_{k'l'kl} &=& \int \int dx dy \phi^*_{k'} \phi^*_{l'}[- g_{2D} {\psi^*_{+,0}}\psi_{-,0}] \phi_{k}\phi_{l}\nn\\
K_{k'l'kl} &=& \int \int dx dy \phi^*_{k'} \phi^*_{l'}[\bar{L}_-] \phi_{k}\phi_{l}\nn\\
L_{k'l'kl} &=& \int \int dx dy \phi^*_{k'} \phi^*_{l'}[g_{2D} \psi^2_{-,0}] \phi_{k}\phi_{l}\nn\\
O_{k'l'kl} &=& \int \int dx dy \phi^*_{k'} \phi^*_{l'}[-g_{2D} {\psi^*_{-,0}}^2] \phi_{k}\phi_{l}\nn\\
F_{k'l'kl} &=& -A_{k'l'kl}, \hspace{0.1cm} I_{k'l'kl} = - H_{k'l'kl}, \hspace{0.1cm}
J_{k'l'kl} = D_{k'l'kl}\nn\\
M_{k'l'kl} &=& G_{k'l'kl}, \hspace{0.1cm}
N_{k'l'kl} = -C_{k'l'kl}, \hspace{0.1cm} P_{k'l'kl} = -K_{k'l'kl}\nn
\eea

\section{Details of Derivation from Section \ref{strufac}}
\label{appC}

\subsection{Dynamic Structure factor in Lehmann representation}
The dynamic correlation of a density fluctuation at time t=0 and one at time `t' is given as
\bea
S_D({\bf k},t) &=& \bra{0}\hat{\rho}_{\bf k}(t)\hat{\rho}_{\bf k}(0)\ket{0}\label{df}
\eea
Introducing a complete set of eigenstates of the Hamiltonian $\ket{j}$ and using the time evolution of the density fluctuation i.e.,  $\hat{\rho}_{\bf k}(t) \ket{0}= e^{i\hat{H}_{eff}t}\hat{\rho}_{\bf k}e^{-i\hat{H}_{eff}t}\ket{0}$, Eq.(\ref{df}) becomes
\bea
S_D({\bf k},t) &=& \sum_j \bra{0}e^{i\hat{H}_{eff}t}\hat{\rho}_{\bf k}e^{-i\hat{H}_{eff}t}\ket{j}\bra{j}\hat{\rho}_{\bf k}\ket{0}\nn\\
&=& \sum_j |\bra{0}\hat{\rho}_{\bf k}\ket{j}|^2 e^{-i(\epsilon_j-\epsilon_0)t}\label{df2}
\eea
where, $\hat{H}_{eff}$ is the effective Hamiltonian used for obtaining Eq.(\ref{gpe}) and is given as
\bea
\hat{H}_{eff} &=& \frac{1}{2m} (\hat{p}_x \check{I} - \check{\sigma}_z m \eta)^2 + \frac{\hat{p}_y^2}{2m_y}\check{I} \nn\\
&+& \bigg[V_{Trap} + g_{2D}\bigg( |\psi_+|^2+|\psi_-|^2\bigg)\bigg]\check{I}\label{heff}
\eea

Taking the Fourier transform of Eq.(\ref{df2}), we get
\bea
S_D({\bf k},\omega)&=& \sum_j |\bra{0}\hat{\rho}_{\bf k}\ket{j}|^2 \delta(\omega -(\epsilon_j-\epsilon_0)/\hbar)\\\nn
\eea
The above expression is written in Lehmann representation \cite{Dpines}. Similarly, the DSF for spin-density can be computed.
\subsection{Details of matrix element in Eq.(\ref{dsf})}\label{appC2}
In this part of Appendix, we will compute the matrix element $\bra{0} \hat{\rho}({\bf r}) \ket{j}$. To this purpose, we define the field operators corresponding to each component as,
\bea
\hat{\bm{\psi}}({\bf r}) = e^{\frac{-i\mu t}{\hbar}}\bigg(\hat{\bm{\psi}}_{0}({\bf r})
+ \sum_j \bigg\{
{\bm {u}}_{j}({\bf r}) \hat{b}_j e^{-i\omega_j t} + {\bm{ v}}^*_{j}({\bf r})\hat{b}^\dagger_j e^{i\omega^*_j t}\bigg\}\bigg)\nn\\ \label{foperator}
\eea
where, $\hat{\psi} ({\bf r})= [\hat{\psi}_{+}({\bf r})\hspace{0.2cm}
\hat{\psi}_{-}({\bf r})]^T$, $ \hat{\psi}_{0}({\bf r}) =[\hat{\psi}_{+,0}({\bf r})\hspace{0.2cm}
\hat{\psi}_{-,0}({\bf r})]^T$, ${\bm{ u}}_{j}({\bf r}) = [u_{+,j}({\bf r}) \hspace{0.2cm}
u_{-,j}({\bf r})]^T$ and, ${\bm{ v}}_{j}({\bf r}) = [v_{+,j}({\bf r}) \hspace{0.2cm}
v_{-,j}({\bf r})]^T$.

The density operator corresponding to each component is given as, 
\bea
|\hat{\psi}_+({\bf r})|^2&=&|\hat{\psi}_{+,0}({\bf r})|^2\nn\\
&+&\sum_j \hat{b}_j e^{-i\omega_j t}[u_{+j}(\bf r)\psi^*_{+,0}({\bf r})+v_{+j}(\bf r)\psi_{+,0}({\bf r})]+c.c.\nn\\
|\hat{\psi}_-({\bf r})|^2&=&|\hat{\psi}_{-,0}({\bf r})|^2\nn\\
&+&\sum_j \hat{b}_j e^{-i\omega_j t}[u_{-j}(\bf r)\psi^*_{-,0}({\bf r})+v_{-j}(\bf r)\psi_{-,0}({\bf r})]+c.c.\nn\\
\eea
The matrix element $\bra{0} \hat{\rho}({\bf r}) \ket{j} = \bra{0} [\hat{\rho}({\bf r}),\hat{b}^\dagger_j] \ket{0} $ as $\hat{b}^\dagger_j \ket{0}=\ket{j}$. Therefore, the matrix element corresponding to operators for total density $\hat{\rho}({\bf r}) = |\hat{\psi}_+({\bf r})|^2+|\hat{\psi}_-({\bf r})|^2$ and spin-density $\hat{s}_z({\bf r}) = |\hat{\psi}_+({\bf r})|^2-|\hat{\psi}_-({\bf r})|^2$ can be expressed as,
\bea
\bra{0} \hat{\rho}({\bf r}) \ket{j}&=& e^{-i\omega_j t}\big\{[u_{+j}(\bf r)\psi^*_{+,0}({\bf r})+v_{+j}(\bf r)\psi_{+,0}({\bf r})]\nn\\&+&[u_{-j}(\bf r)\psi^*_{-,0}({\bf r})+v_{-j}(\bf r)\psi_{-,0}({\bf r})]\big\}\nn\\
\bra{0} \hat{s}_z({\bf r}) \ket{j}&=&e^{-i\omega_j t}\big\{[u_{+j}(\bf r)\psi^*_{+,0}({\bf r})+v_{+j}(\bf r)\psi_{+,0}({\bf r})]\nn\\&-&[u_{-j}(\bf r)\psi^*_{-,0}({\bf r})+v_{-j}(\bf r)\psi_{-,0}({\bf r})]\big\}\nn
\eea

\end{document}